\documentclass[fleqn,usenatbib]{mnras}

\usepackage{newtxtext,newtxmath}

\usepackage[T1]{fontenc}
\usepackage{ae,aecompl}

\newcommand{\Gaiaurl}{\href{https://gea.esac.esa.int/archive/documentation/GDR2/Gaia_archive/chap_datamodel/sec_dm_main_tables/ssec_dm_gaia_source.html}
{\url{https://gea.esac.esa.int/archive/documentation/GDR2/}}}

\usepackage{graphicx}	
\usepackage{amsmath}	
\usepackage{amssymb}	

\PassOptionsToPackage{pdfpagelabels=false}{hyperref}
\PassOptionsToPackage{nonamebreak}{natbib}   

\title[Wide binaries in Gaia]{Imprints of white dwarf recoil in the separation distribution of Gaia wide binaries}

\author[El-Badry \& Rix]{
Kareem El-Badry$^{1}$\thanks{E-mail: kelbadry@berkeley.edu}
and Hans-Walter Rix$^{2}$
\\
$^{1}$Department of Astronomy and Theoretical Astrophysics Center, University of California Berkeley, Berkeley, CA 94720\\
$^{2}$Max Planck Institute for Astronomy, D-69117 Heidelberg, Germany\\
}

\date{Submitted to MNRAS, July 16, 2018}

\pubyear{2018}

\begin{document}
\label{firstpage}
\pagerange{\pageref{firstpage}--\pageref{lastpage}}
\maketitle

\begin{abstract}
We construct from \textit{Gaia} DR2 an extensive and very pure ($\lesssim 0.2\%$ contamination) catalog of wide binaries containing main-sequence (MS) and white dwarf (WD) components within 200\,pc of the Sun. The public catalog contains, after removal of clusters and resolved higher-order multiples, $>$\,50,000 MS/MS, $>$\,3,000 WD/MS, and nearly 400 WD/WD binaries with projected separations of $50 \lesssim s/{\rm AU} < 50,000$. Accounting for incompleteness and selection effects, we model the separation distribution of each class of binaries as a broken power-law, revealing marked differences between the three populations. The separation distribution of MS/MS systems is nearly consistent with a single power-law of slope $-1.6$ over at least $500 < s/{\rm AU} < 50,000$, with marginal steepening at $s > 10,000$\,AU. In contrast, the separation distributions of WD/MS and WD/WD binaries show distinct breaks at $\sim$\,3,000\,AU and $\sim$\,1,500\,AU, respectively: they are flatter than the MS/MS distribution at small separations and steeper at large separations. Using binary population synthesis models, we show that these breaks are unlikely to be caused by external factors but can be explained if the WDs incur a kick of $\sim$\,0.75\,km\,s$^{-1}$ during their formation, presumably due to asymmetric mass loss. The data rule out typical kick velocities above 2\,km\,s$^{-1}$. Our results imply that most wide binaries with separations exceeding a few thousand AU become unbound during post-MS evolution.

\end{abstract}


\section{Introduction}
\label{sec:intro}

Wide binaries (semimajor axes $a\gtrsim 100$\,AU) are thought to form when initially unbound pairs of young stars become bound as their birth cluster dissolves \citep{vanAlbada_1968, Moeckel_2010, Kouwenhoven_2010, Moeckel_2011}. The binding energies of wide binaries are  low, comparable to the kinetic energy of a star moving with velocity 1\,km\,s$^{-1}$. Their orbits are therefore fragile, and can be easily disrupted. Such disruption can occur either due to external gravitational perturbations or due to internal kicks originating from one of the component stars. 

Much of the interest in wide binaries to date has stemmed from the prospect of using their separation distribution as a probe of the Galactic gravitation potential \citep[e.g.][]{Bahcall_1985, Chaname_2004, Yoo_2004, Quinn_2009, Monroy_Rodriguez_2014, Correa_2017, Coronado_2018}. Inhomogeneities in the potential due to stars, molecular clouds, black holes, or other dark objects accelerate the disruption of wide binaries \citep{Retterer_1982, Bahcall_1985, Carr_1999, Yoo_2004}. The wide binary separation distribution therefore places constraints on the number and mass distribution of gravitating objects in the disk and halo, and on the dynamical history of the Galaxy. 

Because wide binaries can also potentially be disrupted by internal processes during post-main sequence evolution, the separation distribution of binaries containing one or more stellar remnants contains information about the late stages of stellar evolution. Wide binaries can be easily become unbound or ``ionized'' by $\sim$km\,s$^{-1}$ velocity kicks.
As a consequence, the separation distributions of binaries containing a white dwarf places strong constraints on the occurrence of velocity kicks in dying stars. Such kicks have be theoretically predicted to occur as a result of anisotropic and/or non-adiabatic mass loss, but evidence for their existence is largely circumstantial \citep{Fellhauer_2003, Davis_2008, Izzard_2010}. Constraining their occurrence is one of the primary goals of this work. 

Historically, searches for wide binaries have identified pairs of stars with close angular separations and/or similar proper motions. Because the orbital timescales of wide binaries are too long for changes in positions of the component stars to be observed in real time, such studies rely on probabilistic arguments to identify pairs likely to be bound. 

In the absence of astrometric data, the wide binary population can be studied statistically through the stellar angular two-point correlation function \citep{Bahcall_1981, Garnavich_1988, Gould_1995,  Sesar_2008, Longhitano_2010, Dhital_2015}: although it is not possible to determine conclusively whether a given pair is a gravitationally bound binary or a chance alignment, the chance alignment rate can be estimated for stochastically distributed stars \citep[e.g.][]{Lepine_2007, Sesar_2008, Dhital_2015}, allowing the number of true binaries to be estimated after the expected number of chance alignments at a given angular separation is subtracted. 

Wide binaries can be identified more reliably with the aid of proper motion data, which can eliminate interloper pairs that appear close on the sky but have significantly different projected angular velocities. Astrometric searches for wide binaries have primarily targeted stars with large proper motions \citep{Luyten_1971, Luyten_1979, Luyten_1979b, Luyten_1988, Wasserman_1991, Poveda_1994, Silvestri_2002, Salim_2003, Chaname_2004, Makarov_2008, Dhital_2010} in order to limit contamination from interlopers that most commonly are distant stars with correspondingly small proper motions. 

The purity of a sample of wide binaries can be further improved with the inclusion of parallaxes \citep[e.g.][]{Close_1990, Shaya_2011, Andrews_2017, Hollands_2018} or radial velocities \citep[e.g.][]{Latham_1984, Close_1990, PriceWhelan_2017, Andrews_2018, Andrews_2018a}. Until recently, most wide binary searches that incorporated parallaxes were
restricted to modest samples of stars in the immediate solar neighborhood ($d\lesssim 50\,\rm pc$). However, the recent {\it Gaia} data releases \citep{Gaia_2016, Brown_2018} have dramatically expanded the sample of stars for which precise parallaxes and proper motions are available, making it possible to assemble large statistical samples of binaries with a $\ll 1\,\%$ contamination rate. 

In this paper, we use {\it Gaia} DR2 data to produce a catalog of wide binaries within 200 pc with very high ($\sim$99.8\%) purity. We divide binaries into main sequence--main sequence (MS/MS), white dwarf--main sequence (WD/MS), and white dwarf--white dwarf (WD/WD) pairs and study the separation distributions of each class of binaries separately. This sample is not volume complete. But the completeness as a function of angular separation can be well characterized (Appendix~\ref{sec:sensitivity}), allowing us to correct for observational biases and infer the intrinsic separation distributions. 

Because white dwarf binaries are faint and rare, past studies of the WD/WD binary population have been limited to samples of a few dozen objects \citep[e.g.][]{Greenstein_1986, Sion_1991, Andrews_2012, Toonen_2017}. A few hundred candidate WD/MS binaries have been identified in common proper motion catalogs \citep{Smith_1997, Silvestri_2002, Chaname_2004, Gould_2004}, but a systematic study of their separation distribution has yet to be carried out. {\it Gaia} astrometry makes it possible to assemble a large, homogenous catalog of WD/MS and WD/WD binaries for the first time.

The remainder of this paper is organized as follows. In Section~\ref{sec:methods}, we describe our strategy for selecting candidate binaries. Section~\ref{sec:sep_dist} describes our method for extracting the intrinsic separation distributions, which are then presented in Section~\ref{sec:result}. These data reveal clear differences in the
separation distributions of MS/MS, WD/MS and WD/WD binaries. In Section~\ref{sec:theory}, we compare our results to the predictions of binary population synthesis models and use the separation distribution of binaries containing a white dwarf to constrain the occurrence of kicks during white dwarf birth. We summarize our findings in Section~\ref{sec:conclusions}. Appendices~\ref{sec:sensitivity}--\ref{sec:query} provide more information on the details of our model. The catalog of our high-confidence binaries is described in Appendix~\ref{sec:catalog}.

\section{Methods}
\label{sec:methods}
For each star in {\it Gaia} DR2 within 200 pc that passes our quality cuts (Section~\ref{sec:quality}), we consider as potential binary companions all neighboring stars that pass similar quality cuts and lie within an aperture of projected radius $5\times10^4$\,AU. We further require the two stars in a pair to have parallaxes and proper motions consistent with a gravitationally bound binary, as described in Section~\ref{sec:sample}. Finally, we remove binaries that are members of clusters, moving groups, and higher-order multiples (Section~\ref{sec:no_friends}). 

We limit our search to projected separations $s<5\times 10^4\,\rm AU\approx 0.25\,\rm pc$, largely because the contamination rate from background sources \citep[e.g.][]{Lepine_2007, Andrews_2017} and unbound associations \citep[e.g.][]{Caballero_2009, Oh_2017, Dhital_2015} grows rapidly at larger separations. Disruption from external gravitational perturbations is  expected to have significant effects only at the largest separations in our catalog \citep[$s\gtrsim 0.1\,\rm pc$;][]{Retterer_1982, Close_1990, Weinberg_1987, Jiang_2010}, so works aiming to use wide binaries as dynamical probes typically focus on binaries with larger separations. As we show in Section~\ref{sec:recoil}, modest velocity kicks during white dwarf formation are expected to ionize binaries with substantially smaller separations. 

\subsection{General quality cuts}
\label{sec:quality}
We require that both members of a candidate binary pair have a five-parameter astrometric solution and a successful measurement of the $\rm G_{BP} - G_{RP}$ color. Following \citet{Lindegren_2018} and \citet{Gaia_2018}, we require both stars to have low astrometric excess noise, satisfying $\sqrt{\chi^2/(\nu'-5)}<1.2\times \max(1, \exp(-0.2(G-19.5))$, where $\chi^2$ and $\nu'$ are respectively referred to as \texttt{astrometric\_chi2\_al} and \texttt{astrometric\_n\_good\_obs\_al} in the {\it Gaia} archive. 

We further require the photometry for both stars to be relatively uncontaminated from nearby sources. Because $\rm G_{BP}$ and $\rm G_{RP}$ magnitudes are calculated by integrating over low-resolution spectra, which are more dispersed than the $G$-band point spread function, the degree of contamination from nearby sources can be assessed by comparing the total BP and RP flux to the $G$-band flux \citep{Evans_2018}. We require both stars to satisfy $1.0+0.015({\rm G_{BP}-G_{RP}})^{2} <$ \texttt{phot\_bp\_rp\_excess\_factor} $ < 1.3+0.06({\rm G_{BP}-G_{RP}})^{2}$. 

Finally, we require both stars to have high-SNR photometry: $<2\%$ flux uncertainties in the $G$-band and $<5\%$ ($<10\%$) uncertainties in BP and RP flux, for the primary and companion, respectively. These cuts remove a substantial number of spurious sources that fall in unphysical regions on the color-magnitude diagram, as well as some real sources \citep[see][]{Gaia_2018}.

In particular, for pairs with small parallaxes and proper motions -- distant and slowly moving stars --  it is essentially impossible to conclusively rule out the possibility that a given pair is a chance alignment of unassociated stars. Any search for wide binaries must therefore make compromises between completeness and purity. This work aims to construct a sample with high purity. This is essential for a reliable measurement of the separation distribution at large separations, where true binaries are rare and contaminants can easily dominate.  

\subsection{Initial Search}
\label{sec:sample}

\begin{figure}
\includegraphics[width=\columnwidth]{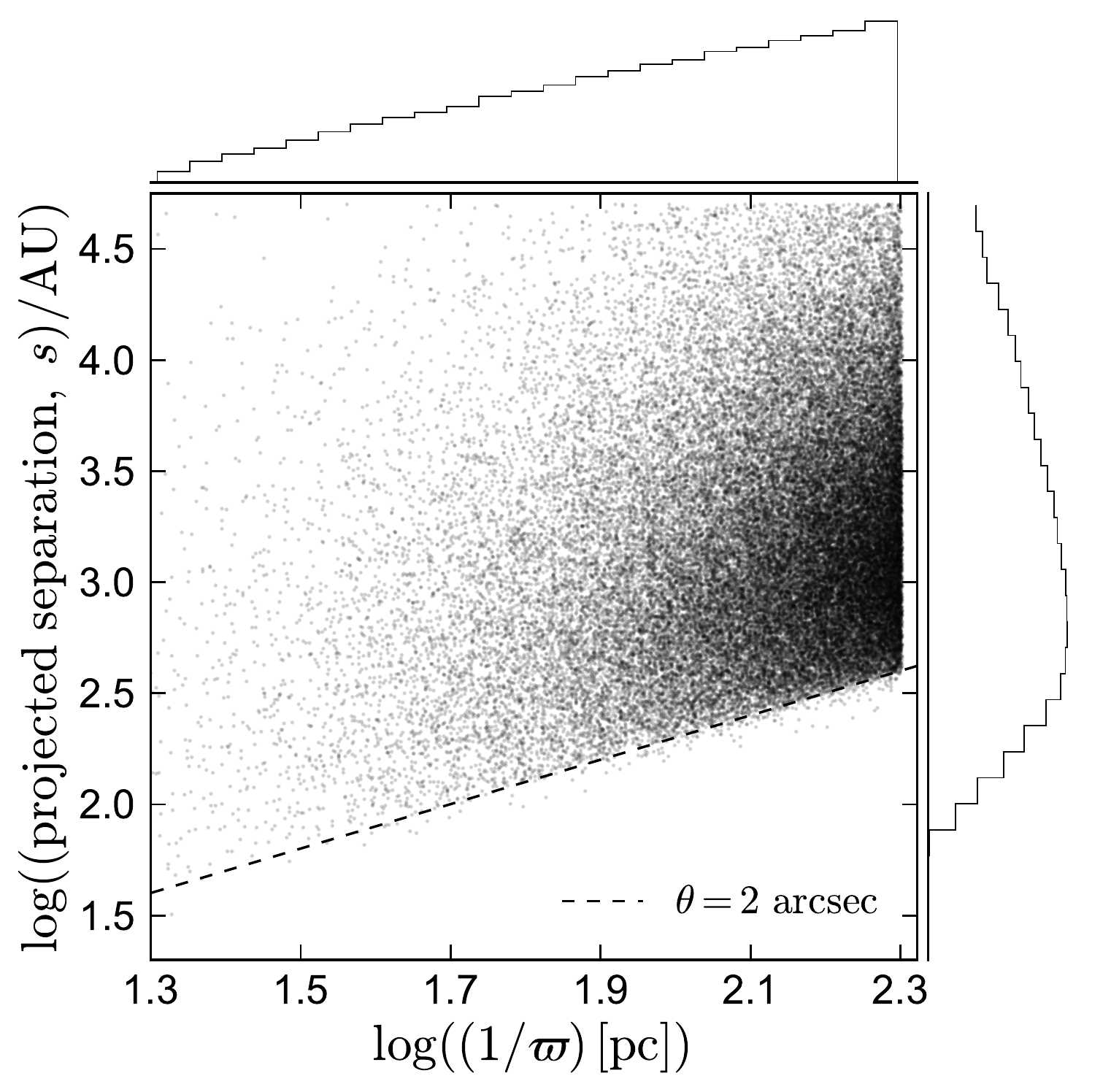}
\caption{Joint distribution of projected physical separation and distance for all 55,128 binaries in our final sample. Marginalized histograms are on a logarithmic scale, with a dynamic range of 100 to 10,000 counts. Given our photometric and astrometric quality cuts (Section~\ref{sec:quality}), the typical angular resolution limit of our sample is $\sim$2 arcsec. This prevents the detection of close binaries at large distances. }
\label{fig:dist}
\end{figure}

\begin{figure}
\includegraphics[width=\columnwidth]{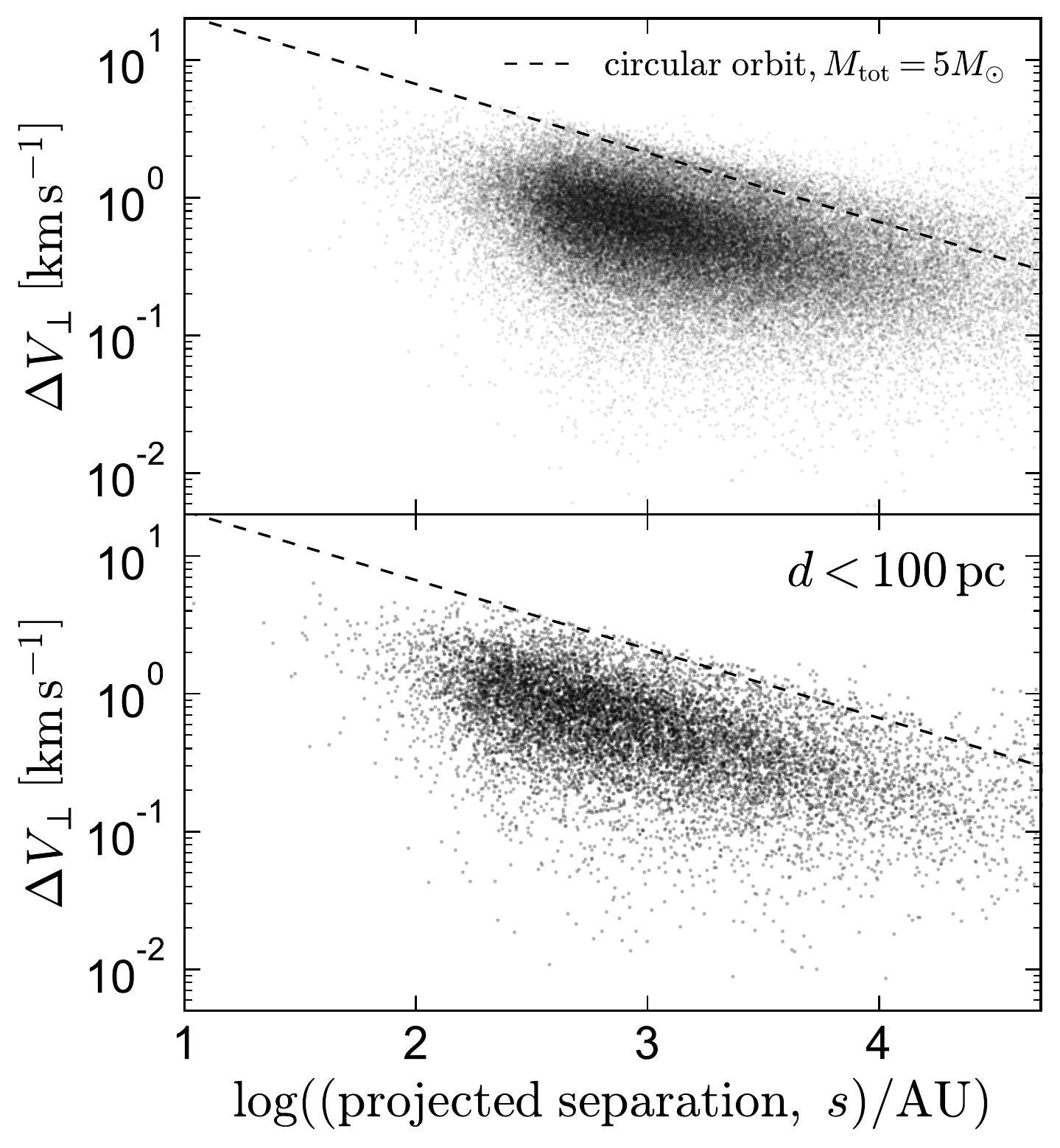}
\caption{On-the-sky physical velocity difference between members of candidate binaries, $\Delta V_{\perp}$, versus their projected physical separation, $s=\theta \times d$. We select binary candidates as objects that fall below the dashed line within $3\sigma$ (Equation~\ref{eq:sigma}); i.e., the difference in their proper motions is small enough to be consistent with a bound Keplerian orbit. Bottom panel shows the subsample with $d<100$\,pc.}
\label{fig:v_perp}
\end{figure}

We search for companions around stars that are nearby and have precise parallaxes, \texttt{parallax} $>5$ and \texttt{parallax\_over\_error} $>20$. We then search a circle corresponding to a projected radius $5\times 10^4$\,AU around each such target, i.e, within an angular separation 
\begin{align}
\frac{\theta}{{\rm arcsec}}\leq 50 \times\frac{\varpi}{{\rm mas}},
\end{align}
where $\varpi$ is the parallax of the target. We also require potential companions to have a reasonably precise parallax, satisfying \texttt{parallax\_over\_error} $>5$. At large distances, one expects the parallaxes of two actual binary components to be essentially identical. However, for the nearest wide binaries, the orbital separation can be a non-negligible fraction of the distance to the binary. Give the high precision of {\it Gaia} parallaxes for nearby stars, this can cause individual stars in a true binary to have inconsistent parallaxes. For most orbits and orientations, the line-of-sight difference in distance between the two stars in a binary satisfies $\Delta d<2 s$. We reject pairs that do not satisfy this constraint within 3$\sigma$, i.e.,

\begin{align}
\Delta d-2s\leq3\sigma_{\Delta d},
\label{eq:dist_constraint}
\end{align}
where in suitable units, $\Delta d=\left|1/\varpi_1-1/\varpi_2\right|$, and $\sigma_{\Delta d}=(\sigma_{\varpi,1}^{2}/\varpi_{1}^{4}+\sigma_{\varpi,2}^{2}/\varpi_{1}^{4})^{1/2}$. Here $\varpi_i$ and $\sigma_{\varpi, i}$ represent the parallax of a target and its standard error. The distribution of sources in our final binary catalog in distance--projected separation space is shown in Figure~\ref{fig:dist}. The {\it Gaia} resolution limit leaves a clear imprint in the distribution, preventing close binaries from being detected at large distances.\footnote{{\it Gaia} DR2 is reasonably complete for separations $\theta \gtrsim 0.7$\,arcsec. However, the majority of pairs with $\theta < 2$\,arcsec lack $\rm G_{BP}- G_{BP}$ colors (see Figure 9 of \citealt{Arenou_2018}) and are therefore excluded from our catalog.} At large separations, the catalog is dominated by the most distant binaries due to the larger volume at large distances. 

We require the difference in proper motion of the two stars to be consistent with a bound Keplerian orbit. For a circular orbit of total mass $5\,M_{\odot}$ and semi-major axis $a$, this translates to a maximum projected physical velocity difference of 
\begin{align}
\frac{\Delta V_{{\rm orbit}}}{{\rm km\,s^{-1}}}\leq 2.1\left(\frac{a}{10^{3}{\rm AU}}\right)^{-1/2}.
\label{eq:delta_v_orbit}
\end{align}
In terms of observables, we translate this to 
\begin{align}
\frac{\Delta \mu_{{\rm orbit}}}{{\rm mas\,yr}^{-1}}\leq0.44\left(\frac{\varpi}{{\rm mas}}\right)^{3/2}\left(\frac{\theta}{{\rm arcsec}}\right)^{-1/2},
\label{eq:orbital_delta_mu}
\end{align}
where we have used the fact that $\Delta V_{\perp}\left[{\rm km\,s^{-1}}\right]=4.74\times\Delta\mu\left[{\rm mas\,yr^{-1}}\right]/\varpi\left[{\rm mas}\right]$.
We do not attempt to account for covariance between parallax and proper motion errors. Equation~\ref{eq:orbital_delta_mu} implicitly assumes that the projected separation $s=\theta\times d$, where $d$ is the distance to a binary, is equal to the true semimajor axis $a$. This assumption does not hold in detail due to projection effects and non-circular orbits. But we show in Appendix~\ref{sec:conversion} that it holds to within a factor of two in a large majority of cases. Equation~\ref{eq:delta_v_orbit} also does not hold in general for non-circular orbits, as the orbital velocity exceeds $\sqrt[]{GM/a}$ near periapse. However, in this case the projected separation is also smaller than $a$, such that Equation~\ref{eq:orbital_delta_mu} holds in the vast majority of cases for randomly oriented orbits \citep[see also][]{Andrews_2017}. 

We require that all candidate binaries have proper motion differences within 3$\sigma$ of the maximum velocity difference expected for a system of total mass $5\,M_{\odot}$ with circular orbits. For each pair, we determine the scalar proper motion difference, 
\begin{align}
\Delta\mu=\left[(\mu_{\alpha,1}^{*}-\mu_{\alpha,2}^{*})^{2}+(\mu_{\delta,1}-\mu_{\delta,2})^{2}\right]^{1/2},
\label{eq:delta_mu}
\end{align}
where $\mu_{\alpha,i}^{*}\equiv \mu_{\alpha,i}\cos\delta_{i}$. The corresponding uncertainty $\sigma_{\Delta \mu}$ is given by standard error propagation:
\begin{align}
\sigma_{\Delta\mu}=\frac{1}{\Delta\mu}\left[\left(\sigma_{\mu_{\alpha,1}^{*}}^{2}+\sigma_{\mu_{\alpha,2}^{*}}^{2}\right)\Delta\mu_{\alpha}^{2}+\left(\sigma_{\mu_{\delta,1}}^{2}+\sigma_{\mu_{\delta,2}}^{2}\right)\Delta\mu_{\delta}^{2}\right]^{1/2},
\end{align}
where $\Delta\mu_{\alpha}^{2}=(\mu_{\alpha,1}^{*}-\mu_{\alpha,2}^{*})^{2}$ and $\Delta\mu_{\delta}^{2}=(\mu_{\delta,1}-\mu_{\delta,2})^{2}$. We reject pairs with $\sigma_{\Delta\mu}\geq 1.5\,{\rm mas\,yr^{-1}}$ .
We then require 
\begin{align}
\Delta\mu\leq\Delta\mu_{{\rm orbit}}+3\sigma_{\Delta\mu},
\label{eq:sigma}
\end{align}
where $\Delta \mu_{\rm orbit}$ is given by Equation~\ref{eq:orbital_delta_mu}. 

The results of this selection are shown in Figure~\ref{fig:v_perp}. The dashed line shows the fiducial cut from Equation~\ref{eq:orbital_delta_mu}. Pairs with significant proper motion uncertainties scatter above it; this scatter is smaller in the 100 pc sample, where typical proper motion and parallax uncertainties are smaller. To limit contamination from pairs with large uncertainties, we impose an absolute cut of $\Delta \mu \leq 2\times \Delta\mu_{{\rm orbit}}$.

Our selection method compares the on-sky projected proper motions of two stars, rather than the 3d space velocities. At large angular separations, this can potentially lead to a bias in the obtained binary population, because projection effects cause stars with the same 3d space velocity to have different proper motions \citep[e.g.][]{Oh_2017}. We performed simulations to assess the magnitude of any possible biases introduced by this affect. Assuming a uniform spatial distribution of binaries and randomly oriented orbits, we found that within 200 pc, 0.7\% of the widest binaries ($s>10^4$\,AU) are expect to have their projected physical velocity difference $\Delta V_{\perp}$ inflated by more than 0.25\,km\,s$^{-1}$ due to projection effects, such that they run a substantial risk of not passing the proper motion cut of Equation~\ref{eq:sigma}. Only binaries with large angular separations ($\theta \gg 10$ arcmin) are affected; so the effect is even smaller for binaries with smaller separations. Within 100 pc, the same fraction affected is 3\%.

\subsection{Removing clusters, moving groups, and higher-order multiples}
\label{sec:no_friends}

\begin{figure}
\includegraphics[width=\columnwidth]{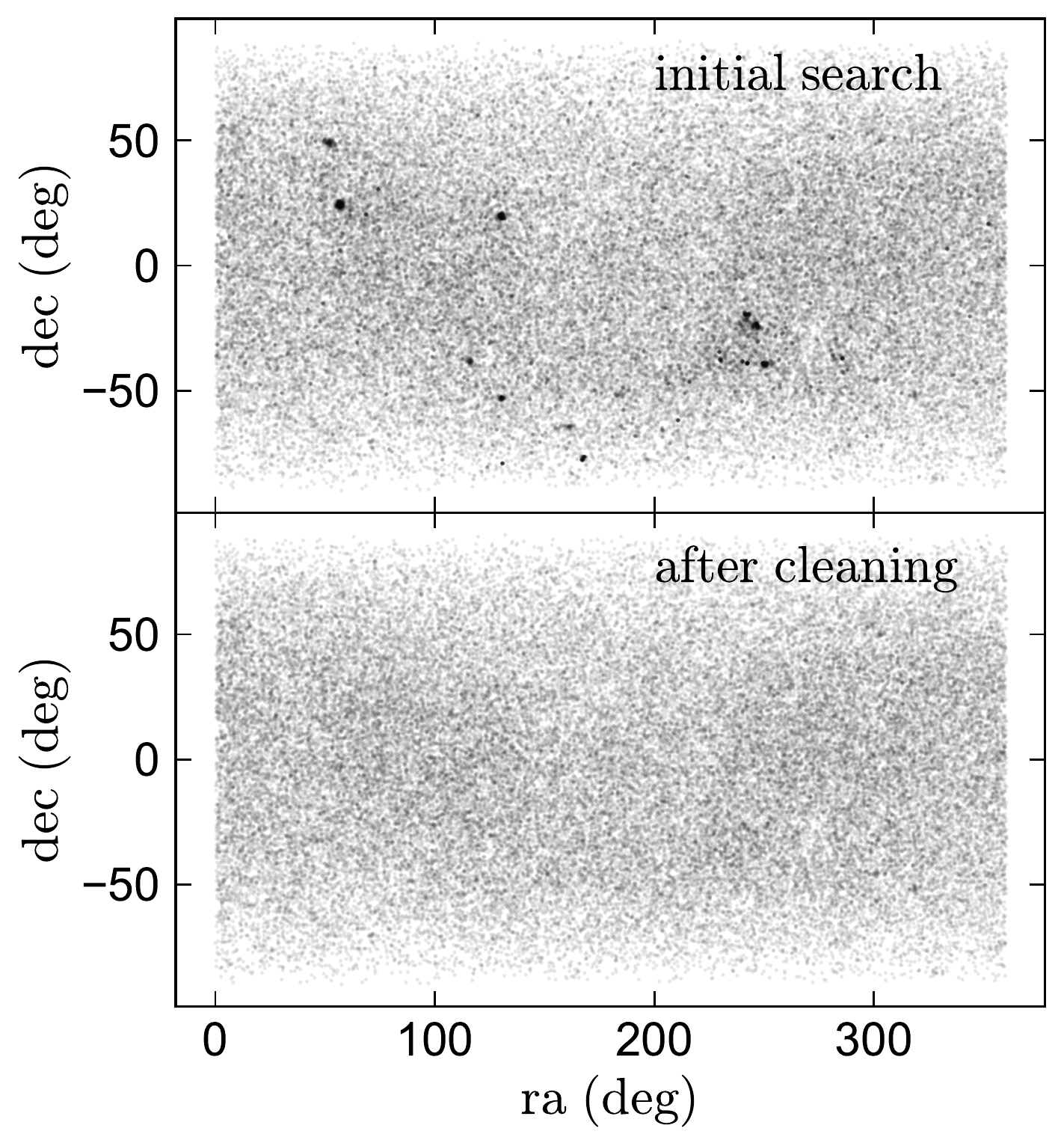}
\caption{{\bf Top}: On-sky distribution of $\sim$60,000 wide, comoving pairs within 200\,pc returned by our initial search. Overdensities correspond to stars in open clusters and moving groups, which meet our initial search criteria but are primarily not genuine bound binaries. {\bf Bottom:} cleaned sample of binaries after $\sim$5,000 pairs in clusters, moving groups, and resolved higher-order multiples have been removed.}
\label{fig:no_friends}
\end{figure}

\begin{figure*}
\includegraphics[width=\textwidth]{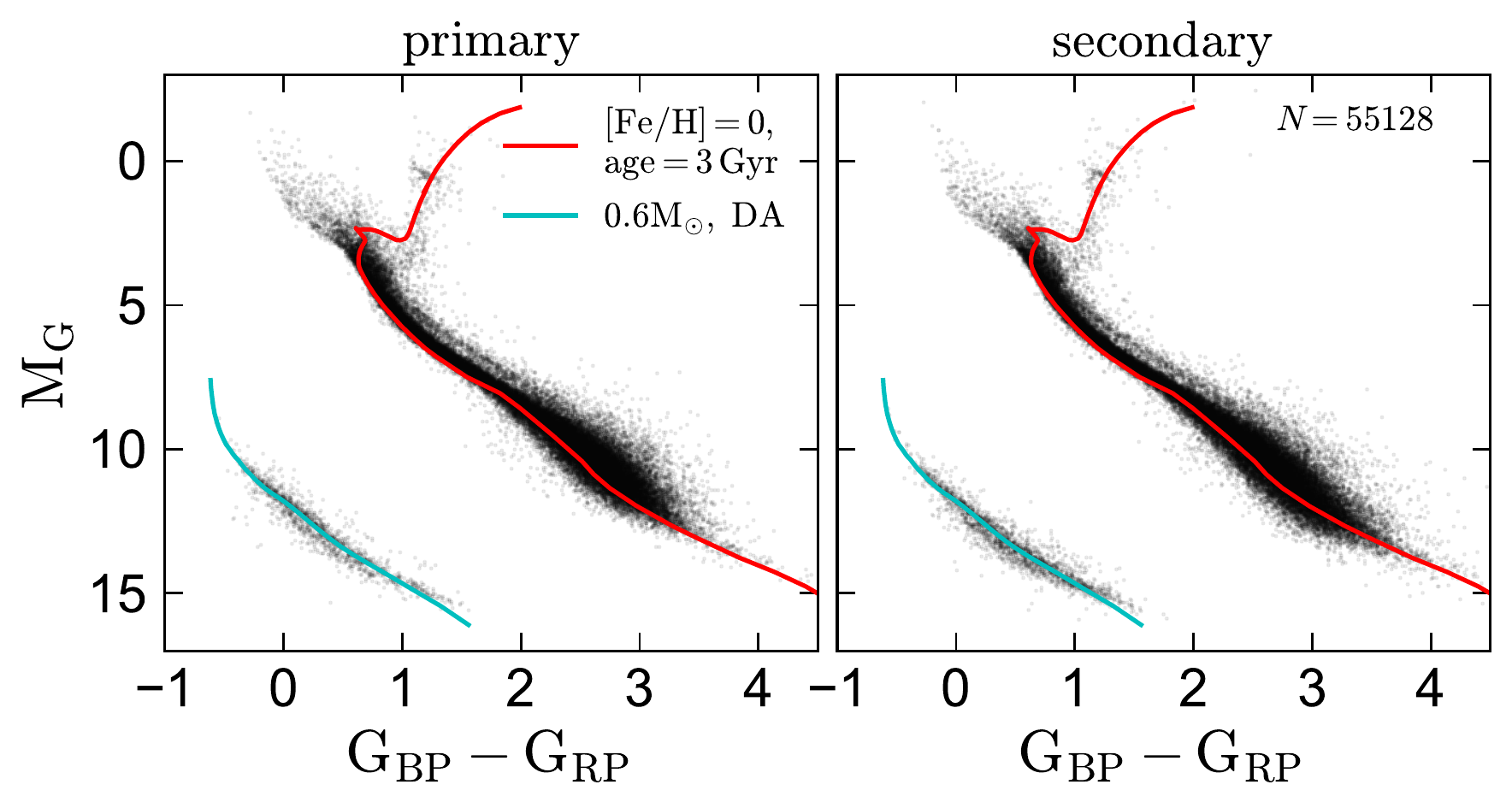}
\caption{Color-magnitude diagram for our final sample of wide binaries. We overplot a solar metallicity PARSEC isochrone and a 0.6\,M$_{\odot}$, hydrogen atmosphere white dwarf cooling sequence. The vast majority of objects in our catalog are main sequence stars and white dwarfs. In each binary, the ``primary'' is the object with the larger photometrically-inferred mass (Section~\ref{sec:summary}).}
\label{fig:cmds}
\end{figure*}

The on-sky distribution of objects returned by our initial search is shown in the upper panel of Figure~\ref{fig:no_friends}. A number of obvious spatial overdensities are apparent. Most of these can be identified with known open  clusters and moving groups within 200 pc. Members of these groups formally satisfy our selection criteria, but most are presumably not genuinely self-bound binaries. Indeed, once-bound associations can remain comoving for several Galactic dynamical times as they slowly dissolve \citep[e.g.][]{Jiang_2010, Shaya_2011, Oelkers_2017, Oh_2017}, so a substantial number of unbound moving groups are predicted to exist. We clean our sample of such objects through two cuts. 

For each candidate binary pair, we count the number of nearby neighbor pairs in position--parallax--proper motion space. We define nearby neighbor pairs as those within 1\,degree on the sky, $\pm 3\,\rm mas\,yr^{-1}$ in both proper motion coordinates, and $\pm 5\,\rm pc$ in $1/\varpi$. We remove from our sample pairs for which this search finds more than 5 neighbors. This cut, which was chosen through trial and error, removes $\sim$2000 candidate binaries, all of which are in one of the overdensities visible in Figure~\ref{fig:no_friends}.

We also repeat the search described in Section~\ref{sec:sample} but replace the proper motion cut from Equation~\ref{eq:orbital_delta_mu} with an absolute cut on $\Delta \mu$ corresponding to a physical velocity difference of 2\,km\,s$^{-1}$. In a majority of cases, this search returns either the same pair found in our initial search, or no results. However, in some cases, it returns many -- in some cases, dozens -- of additional companions to a given object within 50,000 AU. We discard from our sample all pairs for which this search returns more than one companion to either member of a pair. In addition to selecting the majority of cluster and moving group members selected by the first cut, this cut removes $\sim$2,000 triples and resolved higher-order multiples in the field. We defer further analysis of these systems to future work. 

The bottom panel of Figure~\ref{fig:no_friends} shows the effects of these cuts on our sample. Obvious clusters and associations are removed. Although not shown in the figure, we also find that obvious structure in proper motion space is removed. Our final sample contains 55,128 binaries. It does not contain any resolved higher-order multiples; i.e., if a star is a member of more than one pair, we remove both pairs.

\subsection{Summary of the binary catalog}
\label{sec:summary}
A color-magnitude diagram of all objects in our catalog is shown in Figure~\ref{fig:cmds}. We assign the parallax of the brighter star to the companion when calculating absolute magnitudes. We define objects with $\rm M_G < 2.75(G_{BP}-G_{RP})+5.75$ as ``main sequence'' stars\footnote{i.e., our ``main sequence'' category serves only to exclude white dwarfs; it includes a few giants.}, where ${\rm M_{G}=G+5\log\left(\varpi/mas\right)-10}$. Objects with $\rm M_G > 3.25(G_{BP}-G_{RP})+9.63$ are classified as white dwarfs. We do not attempt to correct for extinction, which is minor within 200\,pc. 46 objects that passed our other quality cuts were not classified as either main sequence stars or white dwarfs (i.e., they fell between the main sequence and white dwarf cooling sequence); we discarded the binary candidates containing these objects. Clear sequences of spatially unresolved binaries \citep[e.g.][]{Elbadry_2018b, Widmark_2018} are apparent above the main sequence for both primaries and secondaries, indicating that quite a number of the wide binaries are really hierarchical systems containing unresolved closer binaries. 

Given the color-magnitude diagram, we estimate the mass of each star photometrically by interpolating on a grid of isochrones. We use PARSEC isochrones \citep{Bressan_2012} for main sequence stars and cooling tracks for carbon-oxygen cores with hydrogen atmospheres \citep{Holberg_2006, Kowalski_2006, Tremblay_2011, Bergeron_2011}, in both cases using the revised DR2 filters. These mass estimates are crude and are not included in the public catalog, as we do not make any attempt to correct for unresolved binarity, extinction, or white dwarf spectral types, but in most cases they are expected to be accurate within 0.1 M$_{\odot}$, which is sufficient for characterizing the catalog's binary population in broad strokes. 

\begin{figure*}
\includegraphics[width=\textwidth]{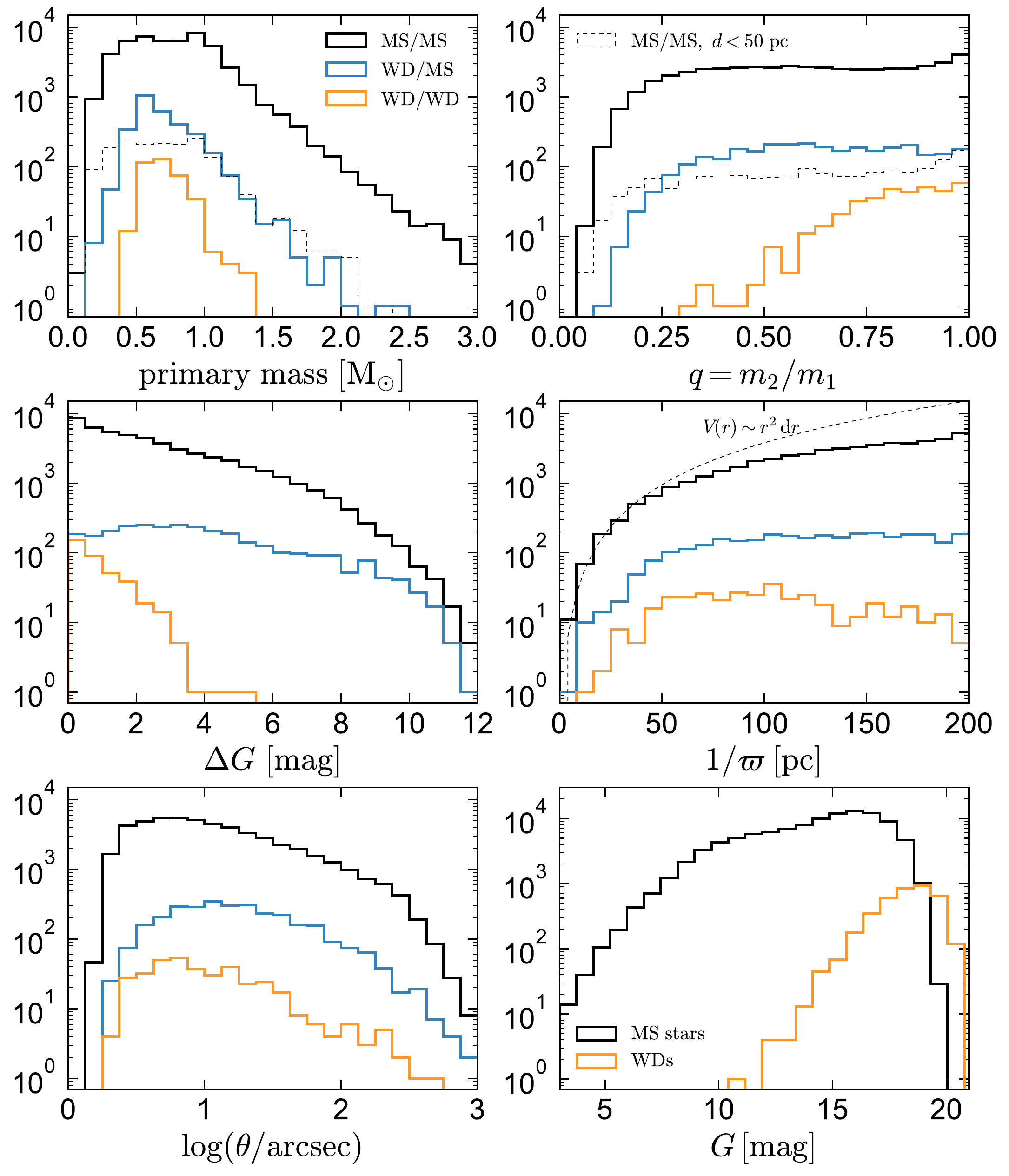}
\caption{Distributions of primary mass, mass ratio, magnitude difference, distance, angular separation, and apparent magnitude for the 51,668 MS/MS, 3,085 WD/MS, and 375 WD/WD wide binaries in our catalog. Most of the main-sequence stars have masses $0.3 \lesssim M/M_{\odot} \lesssim 1.3$; most of the white dwarfs have $0.5 \lesssim M/M_{\odot} \lesssim 0.8$. The mass ratio distribution is roughly uniform for MS/MS binaries. The typical brightness contrast between the two stars in a binary is largest for MS/MS binaries and smallest for WD/WD binaries. Dashed line in the middle right panel shows the scaling expected for a uniform spatial distribution of binaries, illustrating the effects of incompleteness at large distances. The typical angular separation is $\sim$ 10 arcsec, with the widest binaries in the catalog having separations of tens of arcminutes. }
\label{fig:summary}
\end{figure*}

\begin{figure*}
\includegraphics[width=\textwidth]{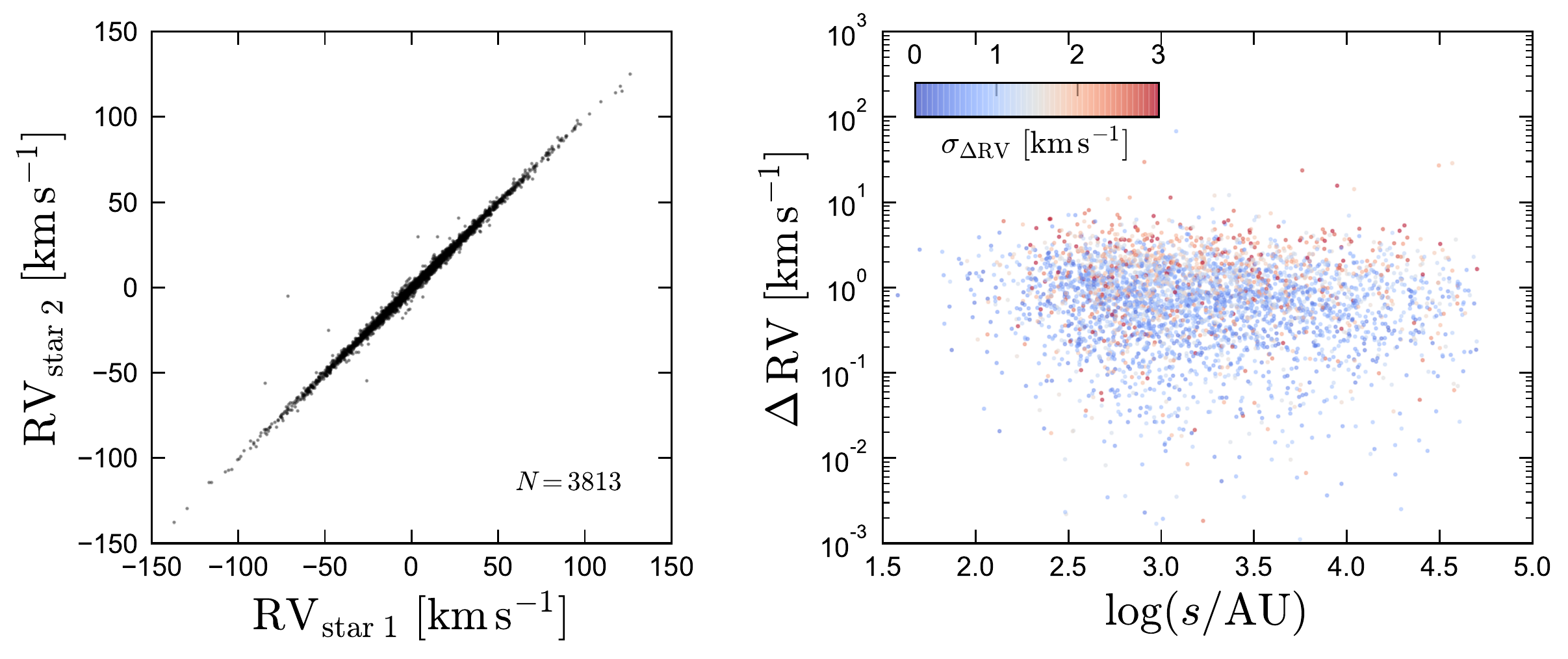}
\caption{{\bf Left}: Heliocentric velocities for the 7.5\% of MS/MS binaries in our catalog for which {\it Gaia} provides a precise radial velocity for both components. Radial velocities were not used in our selection of candidate binaries but provide a useful check on the contamination rate: in genuine binaries, the RVs of the two stars should agree within a few km\,s$^{-1}$ (modulo radial velocity uncertainties of up to 3\,km\,s$^{-1}$). {\bf Right}: Most binaries in our catalog with substantially different radial velocities have larger-than-average radial velocity errors; the fraction of binaries with inconsistent radial velocities does not vary much with orbital separation. We estimate the contamination rate to be $\lesssim 0.2\%$.}
\label{fig:rvs}
\end{figure*}

In Figure~\ref{fig:summary}, we show distributions of several observables and inferred properties for each class of binaries. The top panels show the mass of the primary (which we always define to be the more massive star) and the mass ratio. For white dwarfs, the mass refers to the current mass (as opposed to the initial mass), so in some cases, the white dwarf in a WD/MS binary is the secondary. Most main sequence stars in the sample have masses $0.3 \lesssim M/M_{\odot} \lesssim 1.3$; most white dwarfs have $0.5 \lesssim M/M_{\odot} \lesssim 0.8$. The mass ratio distribution for MS/MS binaries is nearly uniform over $0.25 \lesssim q \lesssim 1$ but exhibits an excess of systems with $q\sim 1$, consistent with previous results \citep[e.g.][]{Soderhjelm_2007}. The dearth of low-mass stars and low-mass ratio binaries can largely be attributed to incompleteness; the fall-off at low masses and mass ratios is less pronounced within 50 pc, where the catalog is more complete.

The middle left panel shows the distribution of apparent magnitude difference, $\Delta G = |G_1 - G_2|$, for binaries in each category. In most of the WD/WD binaries, the two stars have similar apparent magnitude (median $\Delta G=0.66\,\rm mag$). The typical magnitude difference is significantly higher in WD/MS binaries (median $\Delta G=3.50\,\rm mag$) and MS/MS binaries (median $\Delta G=2.05\,\rm mag$). This is important for reliable determination of the intrinsic separation distribution, because the {\it Gaia} sensitivity to a companion at fixed angular separation varies with magnitude difference (see Appendix~\ref{sec:sensitivity}). Corrections for incompleteness at small separations thus affect the three populations differently.

Figure~\ref{fig:summary} also shows the distributions of distance (middle right) and angular separation (bottom left). Because white dwarfs are intrinsically fainter than most main sequence stars, those in our catalog are on average nearer. The median distances for WD/WD, WD/MS, and MS/MS binaries in our catalog are 102 pc, 127 pc, and 147 pc, respectively. The typical angular separation for all binaries is of order 10 arcsec. Due to incompleteness at small angular separations (Figure~\ref{fig:dist}), there are essentially no binaries with angular separations of less than 1.5 arcsec. A few of the nearest binaries have separations of tens of arcminutes, but these are rare; the 99th percentile separation for all binaries in our catalog is 4 arcmin. 

Our requirement that stars in the catalog have precise parallaxes, proper motions, and photometry causes it to contain primarily relatively bright stars (lower right). The median apparent magnitude of MS stars in the catalog is $G=14.5$; for white dwarfs, it is $G=18.4$. Typical parallax and proper motion uncertainties are consequently also higher for white dwarfs. 

\subsubsection{Higher-order multiples}
\label{sec:higher_order}
Although we have removed resolved higher-order multiples from the catalog, it is still expected to contain a substantial number of hierarchical systems in which one or both of the resolved components is really a spatially unresolved close binary. Unresolved binaries with mass ratios $q\gtrsim 0.5$ scatter above the main sequence in a color-magnitude diagram \citep[e.g.][]{Hurley_1998, Widmark_2018}. For solar-type stars, the ``binary sequence'' formed by such objects barely overlaps with the single-star main sequence, making it straightforward to estimate the binary fraction from the fraction of objects above the main sequence.

The unresolved binary sequence is most cleanly separated from the single-star main sequence at $\rm 1 \lesssim (G_{BP} - G_{RP}) \lesssim 2$ (Figure~\ref{fig:cmds}), so we use this region of the CMD to estimate the unresolved binary fraction for MS/MS binaries in our catalog. Main-sequences stars in this region of the CMD have masses $0.5 \lesssim m/M_{\odot} \lesssim 1$. For $\rm 1 < (G_{BP} - G_{RP}) < 2$, the line ${\rm M_{G}=2.8\left(G_{BP}-G_{RP}\right)+2.4}$ divides the binary and single-star main sequences. We find that $\approx$\,10.5\% of primaries and secondaries with $\rm 1 < (G_{BP} - G_{RP}) < 2$ fall above this line and likely have an unresolved companion with $q\gtrsim 0.5$.
Assuming a uniform mass-ratio distribution for unresolved binaries, this implies that $\sim$\,20\% of main-sequence primaries and secondaries have an unresolved main-sequence companion, such that at least one component has an unresolved main-sequence companion in $\sim$\,36\% of the wide binaries in our catalog. Although they cannot be easily identified in an optical CMD, we also estimate that a few percent of the main-sequence components in the catalog have an unresolved white dwarf companion \citep{Willems_2004}.

\subsection{Contamination Rate}
\label{sec:contamination}

Our search for wide binaries in {\it Gaia} DR2 uses only parallaxes, spatial coordinates, and proper motions to determine whether two stars are part of a binary system. We do not use {\it Gaia} radial velocities (RVs) because they are not available for most of our candidates. However, reliable RVs {\it are} available for both stars in 7.5\% of the MS/MS binaries in our sample. We use these to validate our catalog and assess the contamination rate. RVs provide an independent check on whether candidate pairs are true binaries or chance alignments; in the latter case, one would expect the RVs of the two stars to be significantly different, with a typical RV difference comparable to the local velocity dispersion of the disk \citep[e.g.][]{Andrews_2018}. 

We consider all pairs in which a {\it Gaia} RV with uncertainty $\sigma_{\rm RV} < 3\rm \,km\,s^{-1}$ is available for both stars and a {\it Gaia}-RVS spectrum was obtained for at least 3 transits \citep[see][]{Sartoretti_2018, Katz_2018}; $\sigma_{\rm RV}$ is calculated from the scatter in the RVs measured for individual transits. This cut eliminates most objects with bad RVs and yields 3813 pairs. All pairs with RVs are MS/MS, as DR2 does not include RVs for white dwarfs. In the left panel of Figure~\ref{fig:rvs}, we compare the RVs of the two stars in each binary. As expected for true wide binaries, the vast majority of points lie within a few km\,s$^{-1}$ of the one-to-one line. A few binaries do scatter far from the one-to-one line; these are potential contaminants. 

In the right panel of Figure~\ref{fig:rvs}, we show the absolute RV difference, $\Delta \rm RV$, as a function of physical separation and the uncertainty in the RV difference. The majority of binaries with larger-than-average $\Delta \rm RV$ also have larger-than-average $\sigma_{\Delta \rm RV}$, suggesting that they are true binaries. There are 12 binaries with RV differences in excess of $10\,\rm km\,s^{-1}$. Most of these also have relatively large $\sigma_{\Delta \rm RV}$, such that their velocities are marginally consistent with a bound orbit. 6 binaries have $\Delta \rm RV > 15\,\rm km\,s^{-1}$. We regard these as likely contaminants, assuming no catastrophic failure in the RVs. This translates to a contamination rate of $\sim$0.15\%. We note that one might expect the contamination rate for the full sample to be somewhat higher than that in the subsample of pairs with RVs, because the stars with measured RVs are brighter on average and thus have smaller-than-average parallax and proper motion uncertainties.

We obtain an independent estimate of the contamination rate by calculating the number of random alignment pairs expected to pass our selection criteria. We estimate this by applying the same selection criteria and quality cuts to the mock {\it Gaia} DR2 catalog presented in \citet{Rybizki_2018}, which implements the Besan\c{c}on model of stellar population synthesis \citep{Robin_2003} and populates the Galactic distribution function using \texttt{Galaxia} \citep{Sharma_2011}, assuming a  similar selection function and uncertainty model to {\it Gaia} DR2. The mock catalog does not contain any true binaries, so {\it any} pairs that pass our selection criteria in the mock catalog are chance alignments. It also does not contain any white dwarfs, so we restrict our comparison to the MS/MS pairs. 

Executing the same search on the mock catalog, we find 51 pairs. Given a total sample of $\sim$50,000 MS/MS binaries, this implies a contamination rate of 0.1\%. The majority (85\%) of chance alignments found in the mock catalog have $s > 10^4$\,AU. This is expected for chance alignments simply because there is more area at large separations than at small separations. Intriguingly, the binary candidates in our catalog that we identify as likely contaminants because they have large $\Delta \rm RV$ are not strongly concentrated at large $s$ (Figure~\ref{fig:rvs}). This suggests that some of the pairs with large $\Delta \rm RV$ may in fact be genuine binaries in which one of the objects has an unreliable radial velocity. One possibility is that some of these objects are in fact hierarchical triples in which one component is a close, unresolved binary. Most of these systems should be eliminated by the cut of $\sigma_{\rm RV}<3\,\rm km\,s^{-1}$, but a few could escape detection if the different transits occurred at a similar orbital phase. In any case, these tests imply that our sample retains high purity at large separations; for example, the contaminant population estimated from the \citet{Rybizki_2018} mock catalog implies a contamination rate of less than 1\% even at $s > 10^4$\,AU.

\section{Separation distributions}
\label{sec:sep_dist}

\begin{figure}
\includegraphics[width=\columnwidth]{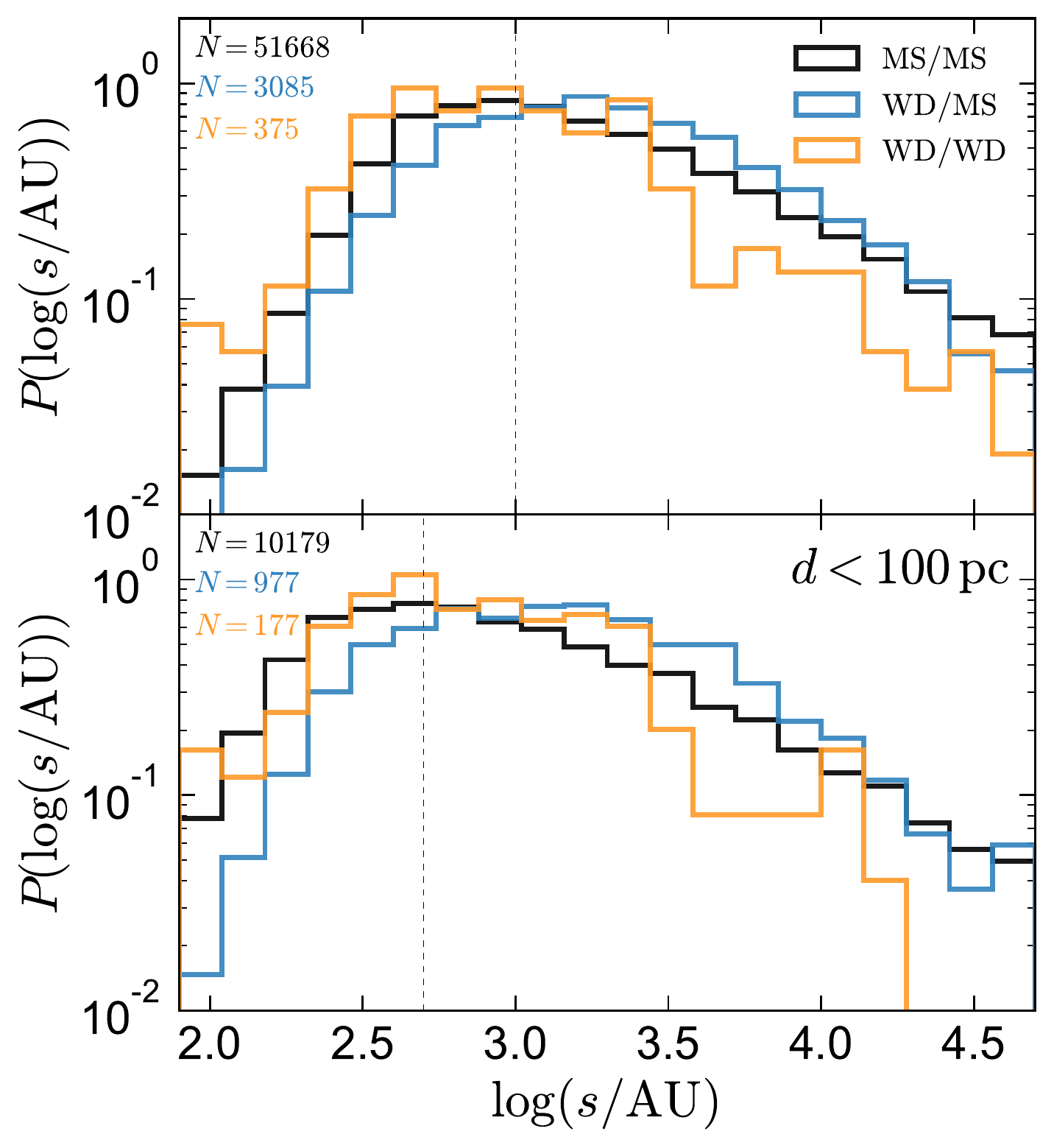}
\caption{Normalized distributions of projected physical separation for MS/MS binaries (black), WD/MS binaries (blue) and WD/WD binaries (gold). Top panel shows full sample within 200 pc; bottom panel shows objects within 100 pc. The dearth of systems with small separations reflects the fact that binaries with small angular separations are spatially unresolved, leading to incompleteness in our sample at $\log(s/\rm AU) \lesssim 3$ in the top panel and $\log(s/\rm AU) \lesssim 2.7$ in the bottom panel (dashed lines; see Appendix~\ref{sec:sensitivity}). The separation distribution of MS/MS binaries falls off as ${\rm d}N/{\rm d}s\sim s^{-1.6}$ (where Opik's law is  ${\rm d}N/{\rm d}s\sim s^{-1}$), at least for $s\gtrsim 500$\,AU. At large separations, the separation distributions of WD/WD and WD/MS binaries fall off more sharply than that of MS/MS binaries. }
\label{fig:sep_dist}
\end{figure}

Figure~\ref{fig:sep_dist} shows raw distributions of the projected physical separation, $s=\theta \times (1/\varpi)$, for the three classes of binaries. The scarcity of binaries with small physical separations is due to incompleteness at small angular separations, where binaries are unresolved. At larger $s$, all binaries are resolved, so the slope of the raw separation distribution is representative of that of the intrinsic distribution. As we show in Appendix~\ref{sec:sensitivity}, within 200 pc the separation distribution is complete for $\Delta G < 5$ (84\% of all binaries) at $s > 1000$\,AU and is complete for all $\Delta G$ at $s > 2000$\,AU. Considering only the 100 pc sample moves the completeness limit toward smaller separations by a factor of two. 

Incompleteness effects become severe at $s \ll 1000$\,AU (full sample) and $s \ll 500$\,AU (100 pc sample), so we caution against overinterpretation of the raw distributions in this regime. Incompleteness affects the separation distribution of each class of binaries in different ways because the have different characteristic magnitude differences (see Figure~\ref{fig:summary} and  Appendix~\ref{sec:sensitivity}). We describe how incompleteness at small orbital separations can be accounted for in Section~\ref{sec:inference}. 

The raw separation distribution of MS/MS binaries in Figure~\ref{fig:sep_dist} clearly continues to rise toward smaller separations, to at least 500 AU. Although the distribution appears to flatten somewhat at smaller separations, it remains unambiguously steeper than Opik's law (i.e., a flat distribution of $\log(s)$). This result is in conflict with the findings of some previous works \citep[e.g.][]{Poveda_2004, Chaname_2004, Lepine_2007, Sesar_2008}, which have found the binary separation distribution to be consistent with Opik's law out to a break at $\sim$3000\,AU, with steeper fall-off beginning only at larger separations. The existence of such a break has been interpreted theoretically as a consequence of dynamical effects within the clusters in which wide binaries are formed \citep{Gould_2006}. Our distribution does not display any strong break. We deem it unlikely that this discrepancy is the result of unaccounted-for systematics in our catalog, as any incompleteness would be expected to introduce biases against {\it smaller} separations.

\subsection{Inferring intrinsic separation distributions}
\label{sec:inference}

Due to our quality cuts and the {\it Gaia} completeness limit, our catalog is not volume-complete: it is missing both spatially unresolved close binaries, and binaries in which either star is too faint to pass our quality cuts or be detected in the first place. Correcting for incompleteness due to unresolved binaries with small separations is important, as it leads to biases in the separation distribution. On the other hand, incompleteness arising from the fact that faint stars are not detected is only a problem if it preferentially affects stars with small or large separations. 

Whether a companion is detected at fixed angular separation is a function primarily of its apparent magnitude, of the magnitude contrast between the two stars, and of the angular separation of the two stars. At sufficiently large angular separations, a binary will be detected as long as both stars are bright enough to be detected independently. However, at small angular separations, the presence of a bright star makes it harder to detect a faint companion at fixed angular separation \citep[see also][]{Ziegler_2018}. This is especially true given the requirement that both stars be free from contamination from nearby sources (Section~\ref{sec:quality}). 

We characterize the probability of a companion's detection at an angular separation $\theta$ with a function $f_{\Delta G}(\theta)$ which depends on $\Delta G$, the $G$-band magnitude difference between the two stars. We empirically constrain $f_{\Delta G}(\theta)$ as described in Appendix~\ref{sec:sensitivity}. As one might expect, $f_{\Delta G}(\theta)$ drops rapidly from 1 at large angular separations\footnote{I.e., $f_{\Delta G}(\theta)$ is normalized relative to the fraction of companions that would be detected at asymptotically large separations, not the absolute number of companions that exist.} to 0 at small separations; the characteristic angular scale at which the drop occurs increases with $\Delta G$. 

We consider an intrinsic separation distribution with functional form $\phi(s|\vec{m})={\rm d}P/{\rm d} s$, where $\vec{m}$ is a set of free model parameters to be fit. Given a sample of binaries with projected separations $s_i$, the likelihood function is 
\begin{align}
L=p\left(\left\{ s_{i}\right\} |\vec{m}\right)=\prod_{i}p\left(s_{i}|\vec{m}\right),
\end{align}
where $p(s_{i}|\vec{m})$ is the probability of detecting the $i$-th binary given model parameters $\vec{m}$. It can be calculated for each binary as
\begin{align}
p\left(s_{i}|\vec{m}\right)=\frac{\phi\left(s_{i}|\vec{m}\right)}{\int_{s_{{\rm min}}}^{s_{{\rm max}}}\phi\left(s|\vec{m}\right)f_{\Delta G}\left(s|d_{i}\right)\,{\rm d}s}.
\label{eq:p_i}
\end{align}
$p_i$ is proportional to the probability that a binary with distance $d_i$, magnitude difference $\Delta G$, and physical separation $s_i$ is found in the catalog, given an intrinsic separation distribution with parameters $\vec{m}$. $f_{\Delta G}$ is explicitly a function of $\theta$, the angular separation of the two stars (Equation~\ref{eq:fitting_func}); it is given by $f_{\Delta G}\left(\theta\right)=f_{\Delta G}\left(s/d_{i}\right)$. Here $s_{\rm min}$ and $s_{\rm max}$ are the minimum and maximum separations of the distribution from which the observed $s_i$ are drawn, and $\phi(s|\vec{m})$ is normalized such that $\int_{s_{{\rm min}}}^{s_{{\rm max}}}\phi\left(s|\vec{m}\right)\,{\rm d}s=1$. We note that Equation~\ref{eq:p_i} does not account for the observational uncertainties in $s_i$ or $d_i$. These are expected to be small, as all the binaries in our sample have parallax errors smaller than 5\%. 

The denominator in Equation~\ref{eq:p_i} represents the fraction of predicted binaries that could have been detected at a distance $d_i$ and magnitude difference $\Delta G$; it accounts for the fact that at large distances and large $\Delta G$, only binaries with large $s$ can be detected.  We set $s_{\rm max} = 5\times 10^4$\,AU. The choice of $s_{\rm min}$ has no effect on our results, since the integrand in Equation~\ref{eq:p_i} goes to 0 at small separations. We set $s_{\rm min} = 10^{-2}$\,AU.  

Whether a binary is detected also depends on the apparent magnitude of both stars, as systems in which either star is too faint will not be detected. This has no effect on the inferred separation distribution as long as the undetected binaries have the same intrinsic separation distribution -- for a particular class of binaries -- as those that are detected. This implies that the separation distribution for each class of binaries must be independent of both distance and the absolute magnitude of the two stars. That $\phi(s)$ does not vary significantly with distance is reasonable within our 200 pc sample, but the separation distribution {\it is} expected vary somewhat with the mass (and thus, absolute magnitude) of the two stars \citep[e.g.][]{Duchene_2013, Moe_2017}. Our inferred separation distribution should thus be viewed as a marginalization over the mass distributions in our sample, which consists primarily of solar and sub-solar mass stars ($0.3\lesssim M/M_{\odot} \lesssim 1.3$; Figure~\ref{fig:summary}).

\subsection{Fitting the separation distribution}
\label{sec:fitting}
We model the separation distribution as a broken power-law: 
\begin{align}
\phi\left(s\right)&=\phi_{0}\begin{cases}
s^{-\gamma_{1}}, & s\leq s_{{\rm break}}\\
s_{{\rm break}}^{\gamma_{2}-\gamma_{1}}\times s^{-\gamma_{2}}, & s>s_{{\rm break}}
\end{cases},
\end{align}
where $\phi_0$ is a normalization constant. The break separation, $s_{\rm break}$, where the distribution transitions from $\phi(s)\sim s^{-\gamma_1}$ to $\phi(s)\sim s^{-\gamma_2}$, is left free, so we fit for three parameters $\vec{m} = (\gamma_1, \gamma_2, \log(s_{\rm break}/{\rm AU}))$. We experimented with using a more flexible parameterization with more than one break in the power law but found that it did not improve the fit much. 
For a power-law parameterization of $\phi(s)$ and our adopted fitting function for $f_{\Delta G}(\theta)$ (Equation~\ref{eq:fitting_func}), the integral in Equation~\ref{eq:p_i} has an analytic solution in terms of hypergeometric functions. 

We sample the posterior distribution for each class of binaries using \texttt{emcee} \citep{FormanMackey_2013}. We use broad, flat priors on all model parameters. 

\subsection{Fitting results}
\label{sec:result}

\begin{figure}
\includegraphics[width=\columnwidth]{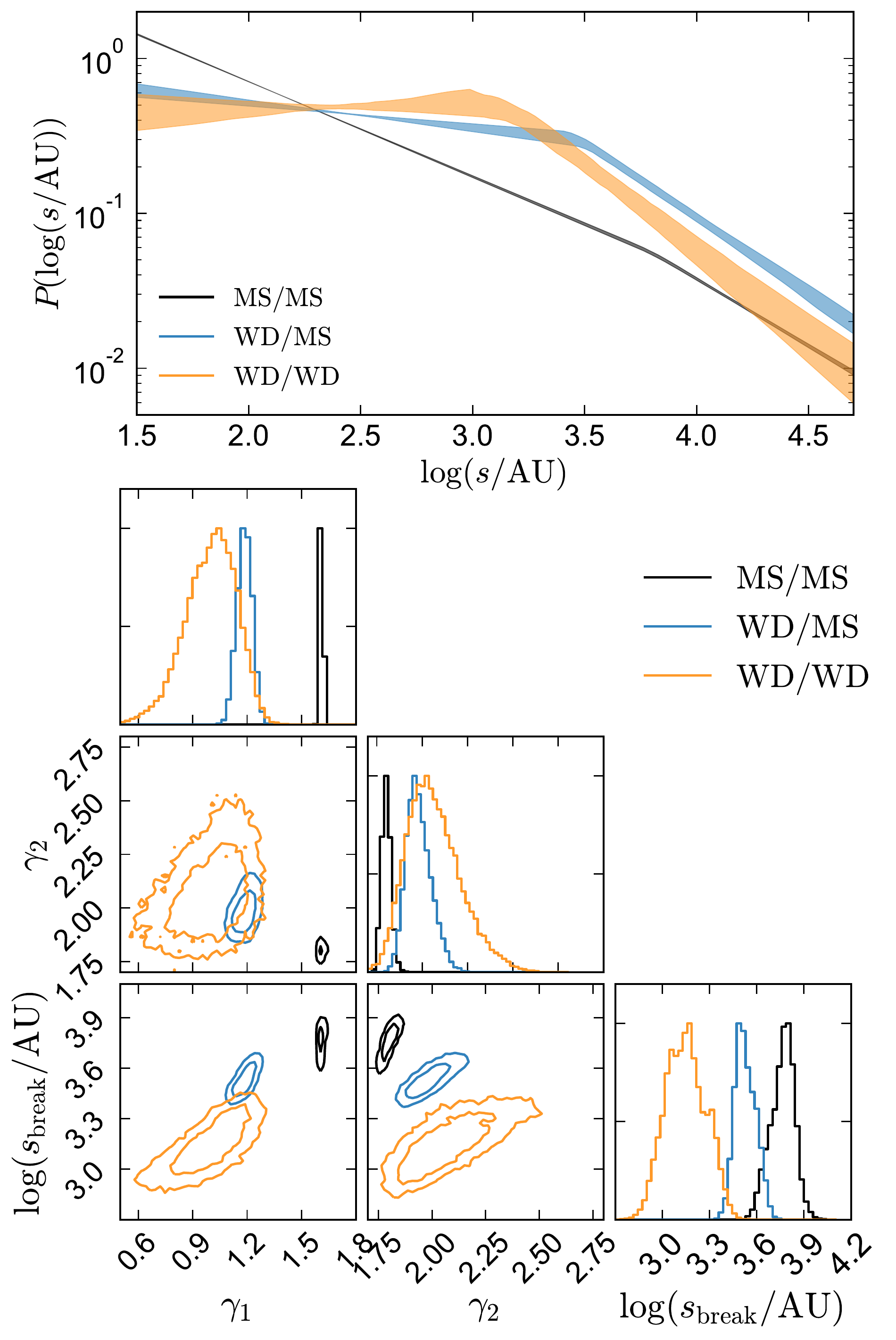}
\caption{{\bf Top}: Broken power-law fits to the intrinsic separation distributions of MS/MS, WD/MS, and WD/WD binaries, after accounting for incompleteness and selection effects. {\bf Bottom}: 68 and 95\% probability contours and marginalized probability distributions for parameters of broken power-law fits. The separation distributions of binaries containing a white dwarf are significantly different from the MS/MS binary separation distribution: they fall off more steeply at large separations and exhibit a stronger break at intermediate separations.}
\label{fig:corner_fits}
\end{figure}

We show the constraints obtained for the separation distribution of each class of binaries in Figure~\ref{fig:corner_fits}. Marginalized parameter constraints are visualized using \texttt{corner} \citep{corner}. Remarkably, these distributions reveal for the first time distinct differences in the separation distributions of wide MS/MS, WD/MS, and WD/WD binaries.   

These differences are already hinted at in the raw distributions, but become more obvious in the inferred intrinsic distributions, which account for the different angular completeness of the three classes of binaries. Consistent with the apparent trends in Figure~\ref{fig:sep_dist}, the separation distribution of WD/WD binaries is almost flat out to $s\sim 1,500$\,AU, after which it drops off sharply. The distribution for WD/MS binaries also exhibits a break, but at slightly larger separations. In contrast, the MS/MS separation distribution is best fit by a model with a weak break, formally at $\log(s/{\rm AU}) = 3.8$. But the change in power law slope is modest (see also the raw distribution in Figure~\ref{fig:sep_dist}), and comparably good fits for MS/MS binaries can be obtained with a single power law model. 

Although there is no strong evidence for a turnover in the separation distributions at small separations in our sample (Figure~\ref{fig:sep_dist}), we emphasize that the true uncertainty in the separation distribution at small $s$ is larger than the shaded regions in the top panel of Figure~\ref{fig:corner_fits}, particularly for MS/MS binaries. As Figures~\ref{fig:dist} and ~\ref{fig:v_perp} show, our catalog contains few binaries with small physical separations. The large number of binaries with separations $2.5 \lesssim \log(s/{\rm AU}) \lesssim 3.5$ set tight constraints on the power-law slope in that range; the slopes of all posterior samples remain tightly constrained at smaller separations because only a single break in the power law is allowed by the model. We find that we obtain comparably good fits to the separation distribution of MS/MS binaries when we fit a lognormal separation distribution that is allowed to flatten at $ \log(s/{\rm AU}) \lesssim 2.5$ (see Appendix~\ref{sec:near_vs_far}). Moreover, when we fit a power law model with an extra break at $\log(s/{\rm AU}) = 2.5$, constraints on the slope at smaller separations become weaker. On the other hand, a steepening in the slope of the separation distributions of WD/MS and WD/WD binaries at intermediate separations is a robust prediction; it is evident in the raw distributions and is recovered for all models that allow a change in the slope of the distribution. 

\begin{figure}
\includegraphics[width=\columnwidth]{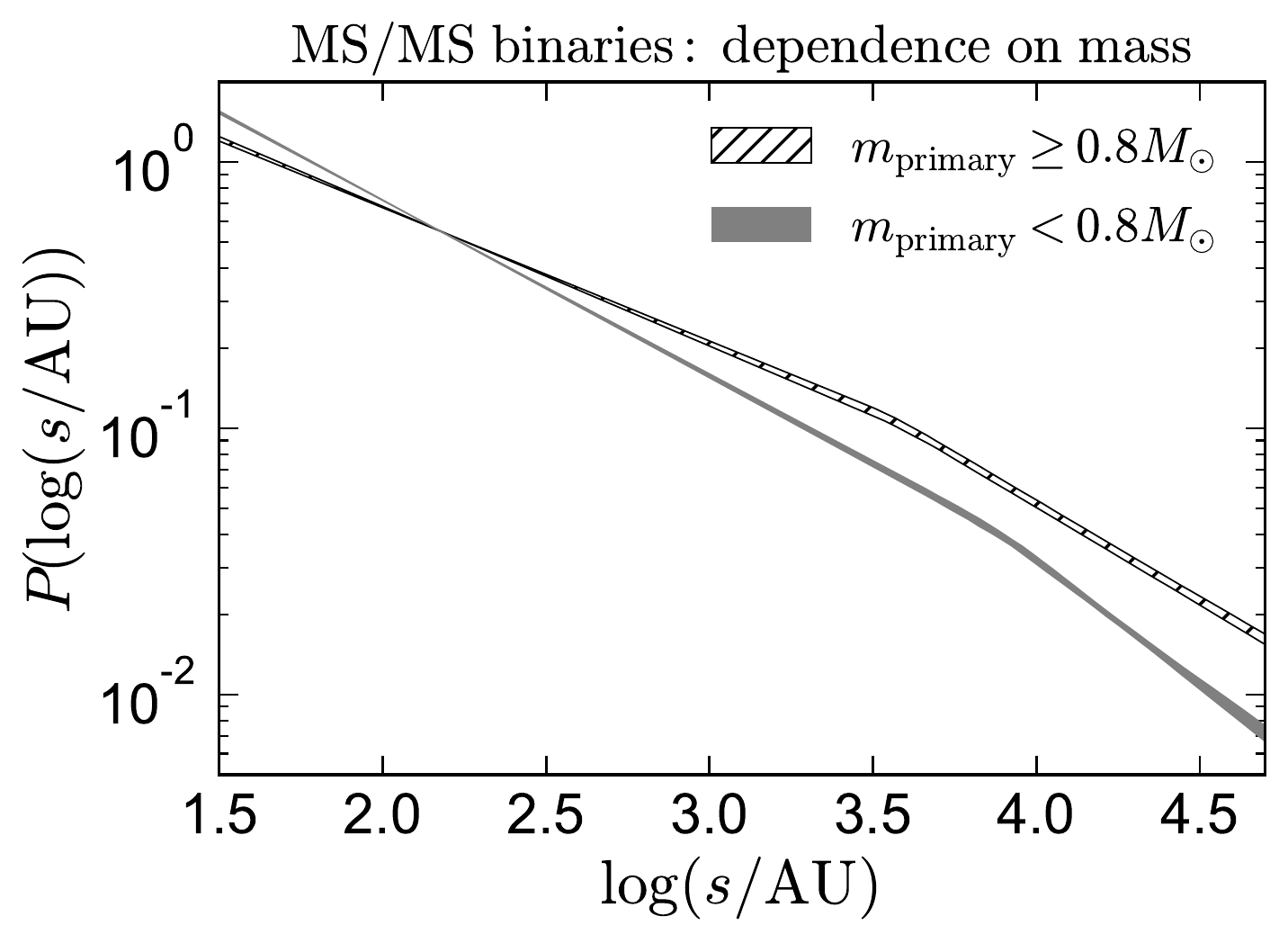}
\caption{Constraints on the separation distribution of MS/MS binaries for systems with higher-mass (hatched) and lower-mass (shaded) primaries. The separation distribution of lower-mass systems falls off more strongly at large separations, consistent with results from previous work.}
\label{fig:mass_dep}
\end{figure}

In Figure~\ref{fig:mass_dep}, we divide the MS/MS sample by mass at a primary mass of 0.8\,$M_{\odot}$ and show separately the inferred separation distributions of MS/MS binaries with higher and lower masses. This cut results in roughly equal numbers of binaries in the two subsamples and is comparable to the minimum mass of a single star that can evolve to become a white dwarf within the age of the Universe.

Consistent with previous work \citep[e.g.][]{Duchene_2013, Moe_2017}, we find that the separation distribution for MS/MS binaries varies somewhat with mass: the distribution for lower-mass systems falls off more strongly at large separations than that for higher-mass systems. Because the progenitors of white dwarfs in our catalog would fall in the higher-mass subsample, variation in the separation distribution with mass should be considered in interpreting the separation distributions of WD/MS and WD/WD binaries. We note that the trend of higher-mass binaries having wider separation distributions is the opposite of what would be expected if the different separation distributions for binaries containing a white dwarf were due only to the higher average masses of white dwarf progenitors.

In Appendix~\ref{sec:near_vs_far}, we verify that our approach yields comparable constraints on the separation distribution when we fit the full catalog and when we fit only the 100 pc sample, which has a different raw separation distribution (Figure~\ref{fig:sep_dist}), completeness, and typical parallax and proper motion uncertainties. This suggest that our inference does a reasonable job of extracting the intrinsic separation distribution without large biases.

\section{Theoretical interpretation}
\label{sec:theory}

Our discovery that the separation distribution of WD/WD (WD/MS) binaries exhibits a break at $\sim$\,1500\,AU ($\sim$\,3000\,AU) that is not seen in the distribution of MS/MS binaries calls for an explanation, which we develop now.

\subsection{Semi-analytic binary evolution models}
\label{sec:bse}

\begin{figure}
\includegraphics[width=\columnwidth]{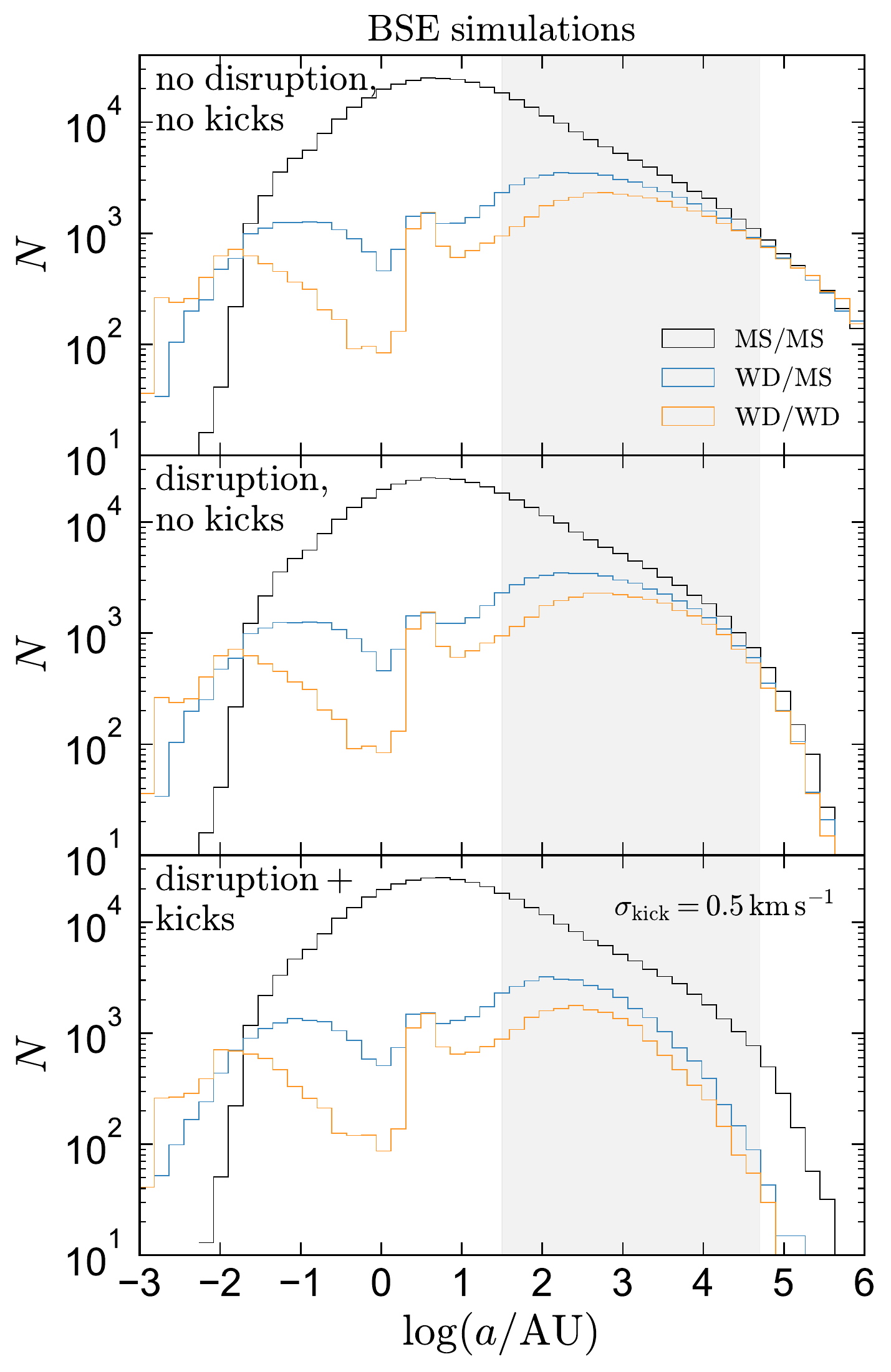}
\caption{Full semi-major axis distributions predicted by our binary population synthesis simulations. Shaded region shows the range of separations probed by our data. Top panel shows the raw population produced by \texttt{BSE}; middle panel shows the results of accounting for the disruption of wide binaries due to external gravitational perturbations (Equation~\ref{eq:disrupt}. In the bottom panel, a modest velocity kick is applied when white dwarfs form due to asymmetric mass loss during post-MS evolution (Section~\ref{sec:recoil}); this unbinds most binaries containing a proto-WD with $\log(a/{\rm AU}) \gtrsim 3.5$ and also expands the orbits of binaries with smaller initial separations.}
\label{fig:bse_population}
\end{figure}

The wide binary separation distribution depends on the intrinsic initial separation distribution, on internal processes that can change the separation of a binary after it forms (e.g. mass transfer, tidal effects, mass loss, or velocity kicks), and on disruption due to gravitational perturbations from other objects. To quantitatively assess the effects of these processes on the observable separation distributions, we carry out Monte Carlo simulations of a large population of synthetic binaries using the semi-analytic model \texttt{BSE} \citep[``binary star evolution'';][]{Hurley_2002}.

We generate $5\times 10^5$ binaries assuming a constant star formation history over 12 Gyr. We draw masses from a \citet{Kroupa_2001b} initial mass function between 0.3\,M$_{\odot}$ and  8\,M$_{\odot}$ and mass ratios from a uniform distribution over [0, 1], roughly consistent with the photometrically-inferred mass distributions for our sample (Figure~\ref{fig:summary}). For systems where the primary mass is greater than 0.75\,M$_{\odot}$, we draw orbital periods from a lognormal distribution with $\overline{\log(P/{\rm day})}=4.8$ and $\sigma_{\log(P/{\rm day})} = 2.3$ \citep{Duquennoy_1991}; for systems with lower mass primaries, we draw from a second lognormal with $\overline{\log(P/{\rm day})}=4.1$ and $\sigma_{\log(P/{\rm day})} = 1.3$ \citep{Fischer_1992}. Eccentricities are drawn from a uniform distribution over [0, 0.95] \citep{Duchene_2013}. We use \texttt{BSE} to evolve all binaries from the zero-age main sequence to present day. We then divide the simulated binary population into MS/MS, WD/MS, WD/WD, and other classes of binaries according to the evolutionary stage reported by \texttt{BSE}. We do not include hierarchical systems containing unresolved binaries in these simulations.

The top panel of Figure~\ref{fig:bse_population} shows the semi-major axis distribution predicted for the three classes of binaries we consider, assuming default model parameters (see Table 3 of \citealt{Hurley_2002}). The distribution for MS/MS binaries is similar to the initial input distribution. The separation distributions of binaries containing a white dwarf are bimodal, with a shortage of systems with $a\sim 1$\,AU. Binaries with initial separations $a\lesssim 10$\,AU interact when their component stars leave the main sequence, and common envelope effects shrink many of their orbits to $a\ll 1$\,AU \citep[e.g.][]{Paczynski_1976, Tout_1991, Ivanova_2013}. 

At larger separations, the components of systems with modest eccentricities do not interact, and the orbital separation {\it grows} when each components leaves the main sequence. In this case -- if mass loss is adiabatic and isotropic -- the orbit gradually expands such that the quantity $(m_1 + m_2)a$ is conserved \citep{Jeans_1924, Rahoma_2009}. This leads the orbits of typical MS/MS binaries to expand by a factor of $\sim$2 on average when one star becomes a white dwarf, and by another factor of $\sim$2 when the second star becomes a white dwarf. These processes make the separation distributions of binaries containing a white dwarf flatter than the MS/MS binary separation distribution in the range of separations probed by our data (top panel of Figure~\ref{fig:bse_population}). 

However, we have not yet accounted for processes that can disrupt the orbits of wide binaries. 

\subsection{Disruption from external perturbations}
\label{sec:disrupt}
All wide binaries can be disrupted over time through gravitational interactions with inhomogeneities in the Galactic gravitational potential caused by stars, molecular clouds, or other dark objects \citep{Heggie_1975, Bahcall_1985, Weinberg_1987, Mallada_2001}. For the most weakly bound binaries, the large-scale Galactic tidal field can also accelerate disruption \citep{Jiang_2010}. 

For the range of separations probed by our data ($a \lesssim 0.25\,\rm pc$), the effects of the Galactic tidal field are negligible. Disruption due to interactions with stars and other objects can be modeled as a diffusive process described by the Fokker-Planck equation. Following \citet{Andrews_2012}, we estimate the timescale $t_{1/2}$ at which a binary of semi-major axis $a$ has a 50\% survival probability as
\begin{align}
t_{1/2}\left(a\right) \approx 29\,{\rm Gyr}\left(\frac{a}{{\rm 10^{4}\,AU}}\right)^{-1}.
\label{eq:disrupt}
\end{align}
This approximation is based on calculations by \citet{Weinberg_1987}. We apply this approximation to our Monte Carlo binary population statistically: the probability of a binary's survival to the present day is set as 
\begin{align}
P_{{\rm survive}}\left(a,{\rm age}\right)=2^{-{\rm age}/t_{{\rm 1/2}}\left(a\right)}.
\label{eq:prob_survive}
\end{align}
In practice, this prescription causes about half of all binaries with $a\sim 5\times 10^4$\,AU to be disrupted. Most binaries with $a>10^5$\,AU are disrupted, while most with $a<10^4$\,AU survive. The middle panel of Figure~\ref{fig:bse_population} shows the effects of disruption on the observable binary population. Disruption affects WD/MS and WD/WD binaries slightly more than MS/MS binaries at fixed semi-major axis because they are older on average, but the difference is minor. This means that the separation distribution of MS/MS binaries can straightforwardly constrain the efficiency of disruption from external perturbations, without potential complications due to the effects of mass loss on binary orbits.

\subsection{White dwarf recoil}
\label{sec:recoil}

We now assess how {\it internal} perturbations arising from kicks to the component stars during post-main sequence evolution are expected to change the observable separation distribution. 

In the calculations described above, \texttt{BSE} assumes that mass loss on the AGB occurs isotropically and quasisteadily over long timescales, such that mass loss leads to gradual orbital expansion. This result will break down if mass loss occurs asymmetrically or too rapidly. If mass loss is asymmetric, the proto-white dwarf will recoil to conserve momentum, resulting in a velocity ``kick'' near the end of the TP-AGB phase or as it forms a planetary nebula. Given typical wind velocities of $(10-20)$\,km\,s$^{-1}$, even 1\% deviations from spherical symmetry in the integrated mass-loss history can produce a recoil velocity of order a km\,s$^{-1}$ \citep{Fellhauer_2003}. Resolved observations of AGB stars find tentative evidence for anisotropic mass loss \citep{Zijlstra_2006, Blasius_2012, Hofner_2018}, and some asymmetric mass loss is likely required to explain the observed rotation rates of single white dwarfs \citep{Spruit_1998}.  

Although the proposed kicks are too weak to detect directly in observations of proto-white dwarfs in planetary nebulae (PNe),\footnote{Within the typical lifetimes of PNe are $\sim 10^4$ yrs \citep{Badenes_2015}, a 1\,km\,s$^{-1}$ kick will transport the WD a distance of order 0.01 pc, which is small compared to the pc-scale size of PNe.} several circumstantial lines of evidence hint at their existence. First, the number of observed white dwarfs in nearby open clusters is lower than is predicted by stellar evolution models for a standard initial mass function, initial mass-age relation, and initial-final mass relation \citep[e.g.][]{Weidemann_1977, Kalirai_2001}; this could be explained if the typical kick speed is comparable to the clusters' few--km\,s$^{-1}$ escape velocity \citep{Fellhauer_2003}. Second, the velocity dispersion of young white dwarfs in globular clusters has been found to be higher than predicted in the absence of natal kicks \citep{Davis_2006, Davis_2008, Calamida_2008}. A typical kick velocity of order 1-2\,km\,s$^{-1}$ could explain these observations \citep{Heyl_2007, Heyl_2008, Heyl_2008b}, and could also play a role in staving off core collapse in globular clusters \citep{Heyl_2009, Fregeau_2009}. Few--km\,s$^{-1}$ velocity kicks during white dwarf formation could also explain the nonzero observed eccentricities of binaries containing a white dwarf and a barium star \citep{Izzard_2010}, as well as the high occurrence rate of dusty disks around young white dwarfs \citep{Stone_2015}.

The separation distributions of WD/MS and WD/WD binaries provide a sensitive probe of white dwarf kicks, as even a small kick is sufficient to unbind a weakly-bound wide binary. For example, for two stars of equal mass $M$ in a circular orbit, the kick velocity required to unbind the system is 
\begin{align}
v_{{\rm kick}}=\left(\frac{GM}{a}\right)^{1/2}=0.5\,{\rm km\,s^{-1}}\left(\frac{M}{M_{\odot}}\right)^{1/2}\left(\frac{a}{10^{3.5}\,{\rm AU}}\right)^{-1/2}.
\end{align}

To test explicitly how kicks during white dwarf formation change the separation distributions of WD/MS and WD/WD binaries, we modify \texttt{BSE} to include a velocity kick when a white dwarf forms. We use the same algorithm \texttt{BSE} uses for (stronger) kicks following SNe explosions. When a white dwarf is formed, we draw a kick velocity magnitude, $v_{\rm kick}$, from a Maxwellian distribution characterized by a free parameter, $\sigma_{\rm kick}$:
\begin{align}
P\left(v_{{\rm kick}}\right)=\sqrt{\frac{2}{\pi}}\frac{v_{{\rm kick}}^{2}}{\sigma_{{\rm kick}}^{3}}\exp\left[-\frac{v_{{\rm kick}}^{2}}{2\sigma_{{\rm kick}}^{2}}\right];
\label{eq:maxwellian}
\end{align}
this distribution peaks at $v_{\rm kick}=\sqrt{2}\sigma_{\rm kick}$. The kick is applied instantaneously in a random direction and occurs at a random time in the orbit. Following a kick, \texttt{BSE} self-consistently solves for the new eccentricity and semi-major axis, as described in Appendix A1 of \citet{Hurley_2002}. If a kick unbinds the system, we consider the binary disrupted and discard it from our subsequent analysis.  

In the bottom panel of Figure~\ref{fig:bse_population}, we show the effects of applying kicks during WD formation with $\sigma_{\rm kick} = 0.5\,\rm km\,s^{-1}$ to the same Monte Carlo synthetic binary population. The main observable effect of including kicks is that the number of very wide WD/MS and WD/WD binaries is reduced, such that their distributions fall off more steeply at $\log(a/{\rm AU})\gtrsim 3$ than the distribution of MS/MS binaries. The disrupted WD/MS and WD/WD binaries with large $a$ are partially replaced by widened binaries with smaller initial separations. But because the orbits of the widest binaries are so weakly bound, only a small number of binaries have initial separations such that their orbits are significantly expanded without being fully disrupted. As a result, even weak kicks with $\sigma_{\rm kick} = 0.5\,\rm km\,s^{-1}$ cause the separation distributions of WD/WD and WD/MS binaries to fall off significantly more steeply at large $a$ than the MS/MS distribution. For $\sigma_{\rm kick} = 0.5\rm \,km\,s^{-1}$, 21\% of MS/WD binaries and 33\% of WD/WD binaries are disrupted. Kicks unbind a majority of WD/MS binaries with initial separations $\log(a_{\rm init}/{\rm AU}) > 3.1$ and a majority of WD/WD binaries with $\log(a_{\rm init}/{\rm AU}) > 2.6$.

\subsubsection{Comparison with observational constraints}

\begin{figure}
\includegraphics[width=\columnwidth]{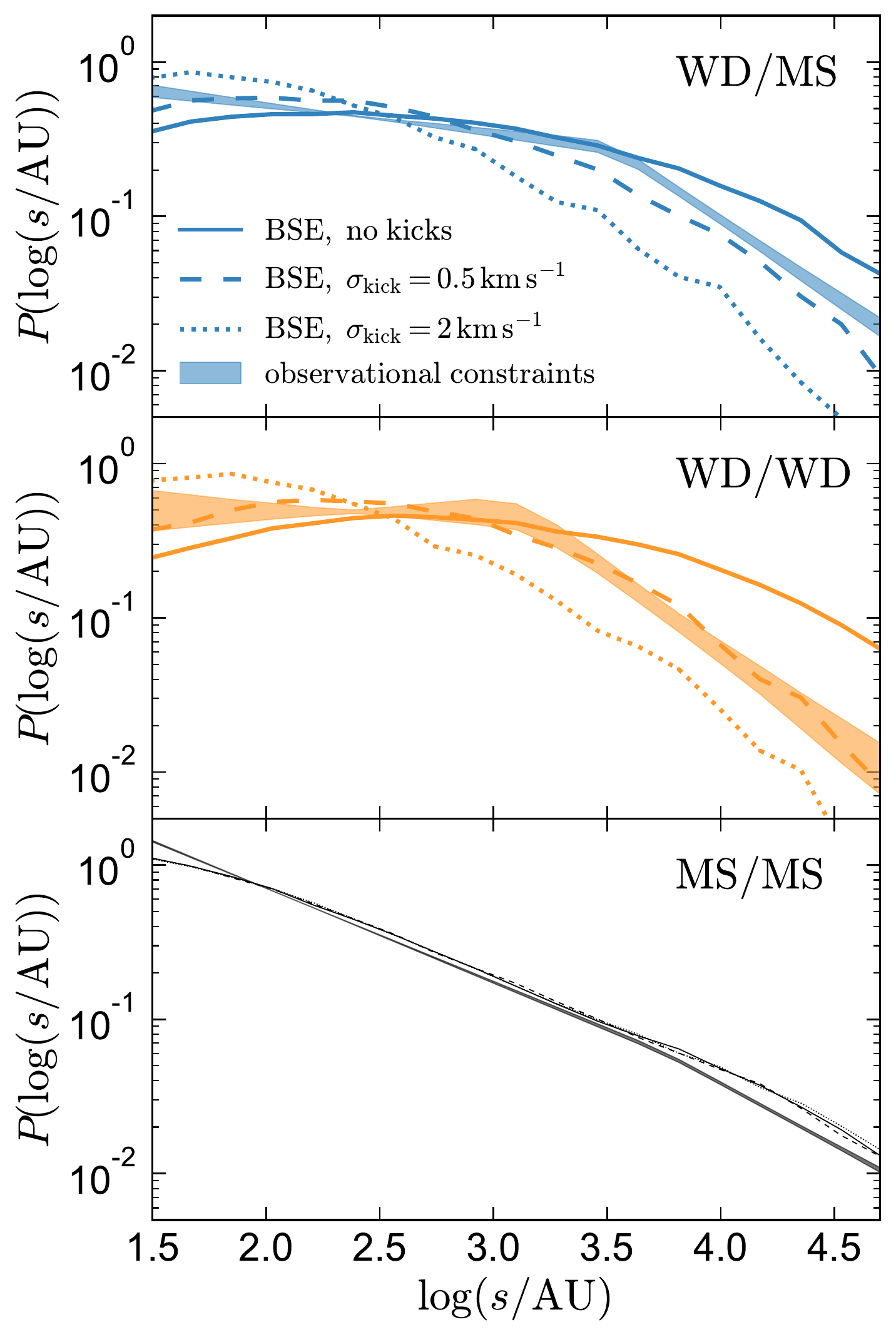}
\caption{Comparison of our separation distribution constraints (shaded regions) to binary population synthesis simulations (``\texttt{BSE}'' ; lines). We compare the results predicted for simulations in which mass loss in post-MS stars is isotropic and adiabatic (``no kicks''; solid lines) to simulations in which a stochastic velocity kick is applied during white dwarf formation (see Section~\ref{sec:recoil}). The separation distributions of WD/MS and WD/WD binaries are best matched by simulations with $\sigma_{\rm kick} = 0.5\,\rm km\,s^{-1}$ (corresponding to a most common kick speed of $\sim$\,0.75\,km\,s$^{-1}$; see Equation~\ref{eq:maxwellian}); simulations with no kicks or larger kick velocities produce substantially worse agreement with the data. The separation distribution of MS/MS binaries (bottom) is independent of kicks but is also reasonably well matched by the simulations. }
\label{fig:kicks}
\end{figure}

We now compare the separation distributions predicted by \texttt{BSE} to our constraints from the {\it Gaia} 200 pc sample. The observational data constrain the distribution of {\it projected} separations, $P(s)$, while \texttt{BSE} returns the distribution of intrinsic orbital semi-major axes, $P(a)$. To carry out a fair comparison between these distributions, we compute in Appendix~\ref{sec:conversion} the probability distribution, $P(a/s)$, of the conversion factor between $s$ and $a$ by mock-observing an ensemble of synthetic binary orbits along random lines of sight. For each binary evolved by \texttt{BSE}, we divide the intrinsic $a$ by a random draw from $P(a/s)$.

The projected separation distributions predicted by \texttt{BSE} for different kick velocities are shown in Figure~\ref{fig:kicks}. Of course, kicks during WD formation have no effect on the separation distribution of MS/MS binaries. The observed constraints on the separation distribution of MS/MS binaries are reasonably similar, though not identical, to the distribution produced by \texttt{BSE}.

The separation distributions of WD/MS and WD/WD binaries appear most consistent with the simulations with $\sigma_{\rm kick} = 0.5\,\rm km\,s^{-1}$, corresponding to a typical kick speed of $\sim$\,0.75\,km\,s$^{-1}$.  Simulations with no kicks produce separation distributions that drop off significantly less steeply than the observational constraints. Conversely, stronger kicks cause the separation distribution of the simulated binary population to drop off more steeply than is observed. This does not rule out the possibility that such kicks occur for {\it some} white dwarfs, but it suggests that they are not widespread. 

\section{Summary and Discussion}
\label{sec:conclusions}

We have constructed a catalog of high-confidence wide binaries using astrometric data from {\it Gaia} DR2. The catalog is publicly available and is described in Appendix~\ref{sec:catalog}. On its basis we determined with high fidelity the physical separation distributions of MS/MS, WD/MS, and WD/WD binaries over the separation range $1.5 \lesssim \log(a/{\rm AU}) \lesssim 4.7$. We discovered that these distributions differ among these types of binaries, and use these differences to constrain the occurrence of kicks during white dwarf formation. We summarize our study below. 

\begin{enumerate}
\item {\it Sample selection}: we require both members of a binary pair to have $d<200$\,pc and have positions, parallaxes, and proper motions consistent with a bound orbit (Figures~\ref{fig:dist} and~\ref{fig:v_perp}). Members of open clusters, moving groups, and wide higher-order multiples are excised from the catalog (Figure~\ref{fig:no_friends}). We classify each component as a main sequence star or a white dwarf using its location on a color-magnitude diagram (Figure~\ref{fig:cmds}). The catalog contains $\sim$50,000 MS/MS binaries, $\sim$3000 WD/MS binaries, and $\sim$400 WD/WD binaries; the main sequence stars are a mix of solar-type stars and M-dwarfs ($(0.3 - 1.3) M_{\odot}$; Figure~\ref{fig:summary}).

\item {\it Purity and completeness}: we estimate the contamination rate of the catalog by comparing the radial velocities of the components of candidate binaries (Figure~\ref{fig:rvs}) and by measuring the chance alignment rate in a mock catalog (Section~\ref{sec:contamination}). Precise parallaxes for both components make it possible to achieve unprecedentedly high purity, with an estimated contamination rate of order $\sim$0.2\%. 

The catalog is not volume complete. Incompleteness at small angular separations due to blending is straightforward to characterize empirically (Appendix~\ref{sec:sensitivity}) and is accounted for in our inference (Section~\ref{sec:inference}). We do not attempt to correct for incompleteness due to missing faint sources, as it does not lead to any obvious biases in the inferred separation distribution. 

\item {\it Separation distributions}: We infer the intrinsic separation distributions for each class of binaries (Figure~\ref{fig:corner_fits}) from the raw distributions (Figure~\ref{fig:sep_dist}). We discovered that the separation distributions of WD/MS and WD/WD binaries fall off more sharply at large separations than the corresponding distribution for MS/MS binaries. The break in the separation distribution occurs at smaller separations for WD/MS binaries than for MS/MS binaries, and at even smaller separations for WD/WD binaries. The log-separation distribution of MS/MS binaries continues to rise toward smaller separations (i.e., it is steeper than Opik's law) to at least $s=500$\,AU (Figure~\ref{fig:sep_dist}). We infer consistent separation distributions for the full sample and the 100 pc sample (Appendix~\ref{sec:near_vs_far}). 

\item {\it Constraints on white dwarf kicks}: We use binary population synthesis simulations to predict the separation distributions of each class of binaries given a reasonable initial separation distribution and dynamical disruption rate (Figure~\ref{fig:bse_population}). These simulations fail to reproduce the observed separation distributions of WD/MS and WD/WD binaries, producing too many binaries with wide separations ($s\gtrsim 5000$\,AU). When modest velocity kicks during white dwarf formation due to asymmetric mass loss are included in the simulations (with typical velocities of $\sim$0.75\,km\,s$^{-1}$), they reproduce the observed separation distributions much more closely (Figure~\ref{fig:kicks}). Ubiquituous kicks with substantially larger kick velocities ($v_{\rm kick}\gtrsim 2$\,km\,s$^{-1}$) can be ruled out, though we cannot exclude the possibility that they occur in a fraction of systems. 
\end{enumerate}

\subsection{Discussion}
\label{sec:discussion}

Introducing a velocity kick during white dwarf formation is of course not the only way to change the predicted separation distribution for binaries containing a white dwarf: one can also change the input initial separation distribution or the prescription for disruption due to external gravitational perturbations, both of which are uncertain. However, the nontrivial differences between the separation distributions of MS/MS binaries and binaries containing a white dwarf (Figure~\ref{fig:corner_fits}) calls for an additional disruption mechanism that affects only binaries containing a white dwarf. 

\citet{Toonen_2017} also considered the possibility that wide binaries are disrupted during white dwarf formation. Rather than modeling explicit natal kicks resulting from anisotropic mass loss, they assumed that mass loss is isotropic but can occur non-adiabatically, on a timescale that is short relative to the orbital timescale. In this case, wide binaries can be unbound when a WD forms because there is not enough time for the orbit to adjust to the reduced binding energy after mass loss \citep[e.g.][]{Savedoff_1966, Veras_2011}. Given a typical AGB lifetime of 1\,Myr, one naively expects such effects to become important for $t_{\rm orbit} \gg 1$\,Myr, corresponding to $a \gg 10^4$\,AU for solar-type stars.

Although \citet{Toonen_2017} did not attempt to model the observed separation distribution, their calculations showed that rapid mass loss could unbind a fraction of white dwarfs in binaries with $a_{\rm init}\gtrsim 5000$\,AU. The primary goal of their simulations was to test whether sudden mass loss could explain the observed deficit of wide WD/WD binaries in the immediate solar neighborhood ($d < 20$\,pc, where the observed white dwarf population is nearly complete) relative to the predictions of populations synthesis models. They found that rapid mass loss would not unbind enough systems to explain the shortage of observed WD/WD binaries, as only a small fraction of all binaries are wide enough to be disrupted. 

We do not attempt to model the normalization of the white dwarf binary population in our sample, because its absolute completeness is not well characterized. Dissolution arising from rapid mass loss like that modeled by \citet{Toonen_2017} could also be partially responsible for the shortage of very wide WD/MS and WD/WD binaries in our sample. However, we note that even if mass loss is instantaneous, it will not necessarily lead to disruption if it is isotropic. For example, a circular orbit will only be disrupted by instantaneous isotropic mass loss if the total mass of both stars decreases by more than 50\% \citep{Huang_1963}, irrespective of the initial separation. This means that equal-mass binaries with circular orbits will {\it always} survive if only one star loses mass. For non-circular orbits, the probability of disruption also depends on orbital phase, but the majority of binaries are expected to survive rapid, isotropic mass loss if the total mass of the binary decreases by less than $\sim$70\% \citep{Savedoff_1966}.

Our population synthesis models assume kicks occur instantaneously; i.e., on timescales that are short compared to the orbital time. For the orbital separations we consider, this is always a good approximation if asymmetric mass loss occurs during PNe formation, but it can become problematic if deviations from spherically symmetric mass loss build gradually, over the full extent of the AGB phase. Using direct N-body simulations of a range binaries orbits, we find that a majority of wide binaries that are disrupted by instantaneous kicks would also be disrupted if the same total velocity change were delivered through a gradual acceleration over many orbital timescales. We thus remain agnostic of the timescale over which anisotropic mass loss is likely to occur. 

Irrespective of the details of the process responsible for ionizing binaries during post-MS evolution, one might also expect the magnitude of kicks to vary with the mass of the white dwarf, as the progenitors of massive white dwarfs lose a larger fraction of their mass during white dwarf birth \citep[e.g.][]{ElBadry_2018}. Splitting the WD/MS binaries in our catalog by photometric white dwarf mass, we do not detect a significant difference between the separation distributions of binaries with lower- and higher-mass white dwarfs. However, the range of white dwarf masses in our catalog is not very large, and the photometric mass estimates are subject to nontrivial uncertainties. Further investigation of any dependence on the mass of the white dwarf and/or main sequence star represents a useful avenue for future work.

Finally, we note that kicks during white dwarf formation could accelerate the rate of WD/WD mergers and collisions in hierarchical triples, with potentially important implications for the SNe Ia rate. Kozai oscillations can drive inner WD/WD binaries in triples to extremely eccentric orbits, leading to a reduced gravitation wave inspiral timescale and/or head-on collisions of the WDs \citep[e.g.][]{Thompson_2011, Katz_2012}. The efficiency of this mechanism depends on the mutual inclination of the inner and outer binaries. Weak kicks during white dwarf formation could rearrange stable hierarchical triples containing an inner WD/WD binary, driving them to higher mutual inclinations where they are susceptible to the Kozai mechanism \citep[e.g.][]{Toonen_2018}. Simulations are needed to assess the efficiency of such kick-driven orbital rearrangement.

\subsubsection{Using the catalog}
\label{sec:use_cat}

Our sample of binaries serves as a potential resource for studies of the Galactic binary population, including the physics of star formation, ISM chemistry, cluster dissolution, and dynamical disruption. We make the catalog of high-confidence binaries within 200 pc publicly available in the hope that others find it useful. 

Although the current catalog is truncated at separations of 50,000 AU $\approx$ 0.25 pc, extending the search to larger separations is straightforward (see Appendix~\ref{sec:query}); doing so may be useful for studies focused on dynamical disruption. At significantly larger separations, a more careful contamination model may be needed to account for unavoidable chance alignments of gravitationally unassociated pairs. 

As discussed in Section~\ref{sec:sample}, the fact that our binary selection uses on-sky projected proper motions can lead to a slight bias against binaries with large angular separations, as stars with identical 3d space velocities can have different proper motions. The effects of this bias are expected to be negligible for our full sample and for the 100 pc sample, but it can lead to a non-trivial selection effect against the widest physical separations if only the nearest binaries are considered. The simulations described in Section~\ref{sec:sample} show that if only binaries within 50 pc (20 pc) are considered, 12\% (60\%) of binaries with $s>10^4$\,AU are likely to be missed. We therefore recommend a less stringent cut on proper motion for studies focused on the nearest binaries. 

For more than 4000 of the binaries in our catalog, at least one component has been observed by a Galactic spectroscopic survey such as LAMOST, RAVE, APOGEE, and GALAH; for several hundred, a spectrum has been obtained for both stars. The metallicities and detailed abundances of binaries in the catalog can be used to investigate the chemical homogeneity of star forming regions, the creation rate through dynamical encounters of wide binaries that did not form together, and the dependence, or lack thereof, of the wide binary fraction on metallicity. 

Wide binaries containing a white dwarf are particularly useful for the determination of stellar ages, as the total age of the white dwarf can be constrained from its cooling age and mass, given an initial-final mass relation and an initial mass-age relation. Assuming that the two components in a binary are co-eval, this age constraint can be transferred to a main-sequence companion, whose age could otherwise not be measured precisely \citep[e.g.][]{Fouesneau_2018}. Such analysis is also useful for calibrating empirical age constraints for main sequences stars based on stellar activity or gyrochronology \citep[e.g.][]{Chaname_2012, GodoyRivera_2018}. The strongest constraints can be obtained for systems in which spectroscopic abundance constraints are available. In additional to the systems which have already been observed by Galactic spectroscopic surveys, we anticipate that {\it Gaia} DR3 will deliver abundances for a large fraction of binaries in our catalog. 

\section*{Acknowledgements}
We are grateful to the referee, Andrei Tokovinin, for a constructive report.
We also thank Carles Badenes, Matteo Cantiello, Andy Casey, Trent Dupuy, Morgan Fouesneau, Andy Gould, Ted von Hippel, David W. Hogg, Sergey Koposov, Chao Liu, Adrian Price-Whelan, Eliot Quataert, Jan Rybizki, David Spergel, and Daniel R. Weisz for helpful discussions. KE was supported by the SFB 881 program (A3) and an NSF graduate research fellowship. We are grateful to Livia DeMarchis for her hospitality during the writing of this manuscript. This project was developed in part at the 2018 NYC Gaia Sprint, hosted by the Center for Computational Astrophysics of the Flatiron Institute in New York City. This work has made use of data from the European Space Agency (ESA) mission {\it Gaia} (\url{https://www.cosmos.esa.int/gaia}), processed by the {\it Gaia} Data Processing and Analysis Consortium (DPAC, \url{https://www.cosmos.esa.int/web/gaia/dpac/consortium}). Funding for the DPAC has been provided by national institutions, in particular the institutions participating in the {\it Gaia} Multilateral Agreement.




\bibliographystyle{mnras}



\appendix

\section{Gaia spatial resolution and dependence on flux ratio}
\label{sec:sensitivity}

Our method of identifying wide binaries requires that the two component stars be spatially resolved and that both stars pass the quality cuts we impose (Section~\ref{sec:quality}). Whether a wide binary is detected thus depends both on the angular separation of the two stars and on their flux ratio. The secondary is more likely to be outshone or contaminated by light from the primary at fixed angular separation if the flux ratio is large than if it is small. This must be accounted for in modeling the intrinsic separation distribution.

\begin{figure}
\includegraphics[width=\columnwidth]{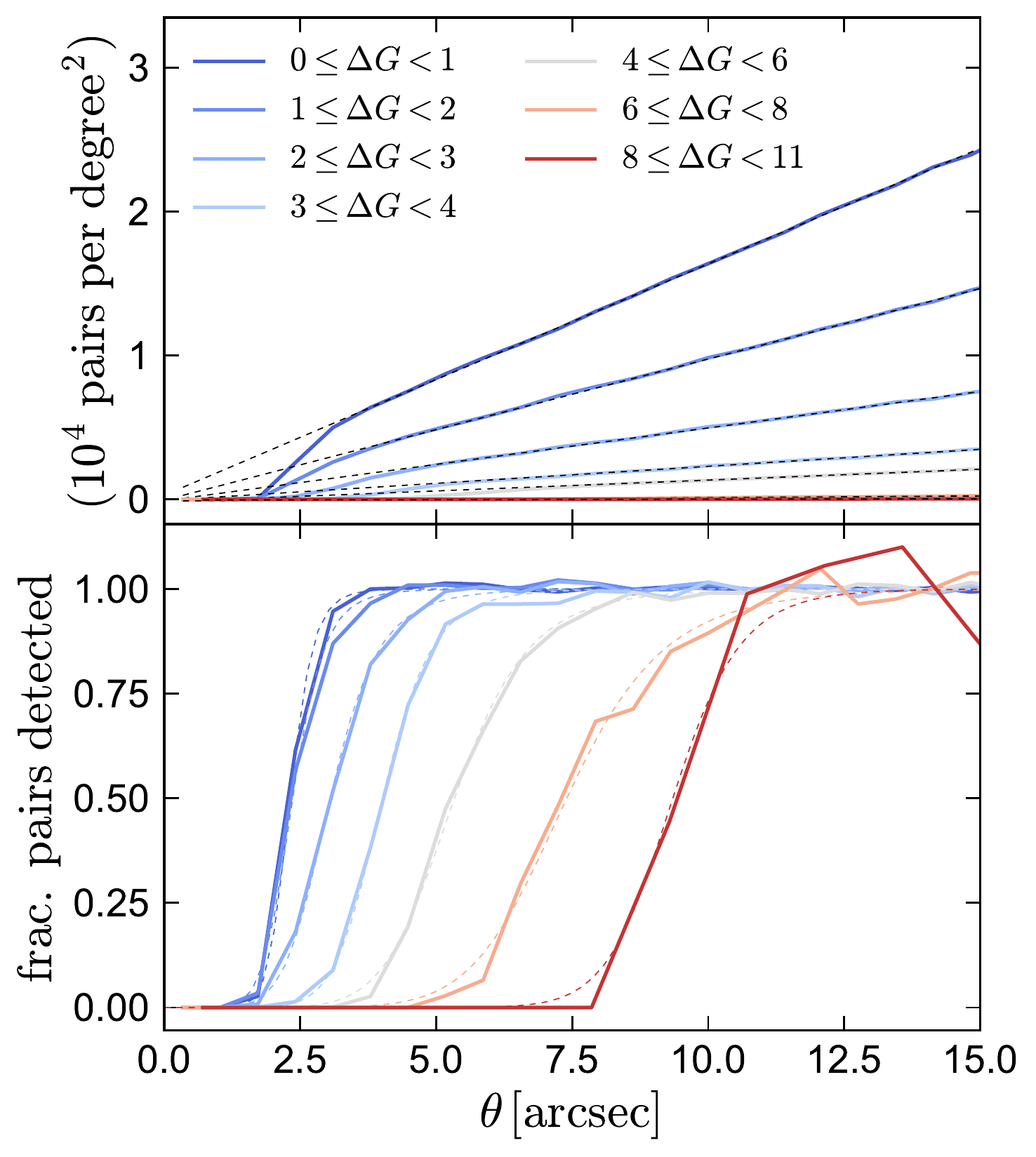}
\caption{{\bf Top}: total number of pairs  with separation $\theta$ (i.e., two-point autocorrelation function) in a dense region ($l, b = -30^{\circ}, -4^{\circ}$). Different lines correspond to pairs in bins of different magnitude difference, $\Delta G=\left|G_1 - G_2\right|$. Dashed lines show linear fits to data at $\theta > 10$\,arcsec. {\bf Bottom}: ratio of data in top panel to linear extrapolations, representing the fraction of all pairs with a given $\Delta G$ that are detected at angular separation $\theta$. Dashed lines show fits using Equation~\ref{eq:fitting_func}.}
\label{fig:sensitivity}
\end{figure}

We assess the sensitivity of {\it Gaia} photometry to a companion at a given angular separation using a method similar to that described in \citet{Arenou_2018}. We query all sources in a field of radius 1.5 degrees and measure the number of pairs passing the quality cuts listed in Section~\ref{sec:quality} as a function of angular separation and $\Delta G$, the $G$-band magnitude difference between the two sources. The resulting two-point correlation function is shown in the top panel Figure~\ref{fig:sensitivity}. For each bin in $\Delta G$, we plot the number of pairs in 0.66 arcsec-wide bins. With no restriction on parallax or proper motion, the vast majority of pairs are chance alignments. At any flux ratio and angular separation $\theta$, the expected number of such random alignments scales as $N_{\rm pairs} \sim 2\pi \theta\,{\rm d}\theta$.

The top panel of Figure~\ref{fig:sensitivity} shows that such a linear scaling is indeed found at large separations for all $\Delta G$ bins. To estimate the total number of pairs at a given $\theta$ and $\Delta G$, including those that are missed due to incompleteness at small separations, we extrapolate a linear fit to the number of pairs found at large separations. We then estimate the fraction of all pairs that are detected by dividing the empirically measured number of pairs by this extrapolation (bottom panel of Figure~\ref{fig:sensitivity}). For all $\Delta G$, the sensitivity drops rapidly from 1 to 0 at some angular separation $\theta_{0}$, where $\theta_0$ increases with $\Delta G$. We fit the sensitivity curves for each $\Delta G$ with an empirical fitting function: 
\begin{align}
\label{eq:fitting_func}
f_{\Delta G}\left(\theta\right)=\frac{1}{1+\left(\theta/\theta_{0}\right)^{-\beta}}.
\end{align}
Here $\theta_0$ characterizes the angular separation below which the sensitivity drops to 0, and $\beta$ determines how rapidly the sensitivity falls off at $\theta \ll \theta_0$. We estimate the value of $\theta_0$ and $\beta$ appropriate for a given $\Delta G$ by interpolating on the fits for the discrete values of $\Delta G$ shown in Figure~\ref{fig:sensitivity}. We find that $\beta = 10$ works reasonably well for all $\Delta G$. For $\theta_0$, we find $\theta_0 \approx 2.25$\,arcsec at $\Delta G < 1.5$\,mag, and $\theta_0 \approx 0.9(\Delta G + 1)$ at $\Delta G > 1.5$\,mag. 

We find that this parameterization of the sensitivity works well over a wide range of colors and source surface densities. However, we caution that it applies only for the particular choice of quality cuts listed in Section~\ref{sec:quality}. \citet{Ziegler_2018} found a comparable sensitivity scaling in the efficiency with which {\it Gaia} recovers known binaries, but with a different normalization $\theta_0$ due to less stringent quality cuts on the {\it Gaia} photometry. 

\section{Relation between projected separation and semimajor axis}
\label{sec:conversion}

\begin{figure}
\includegraphics[width=\columnwidth]{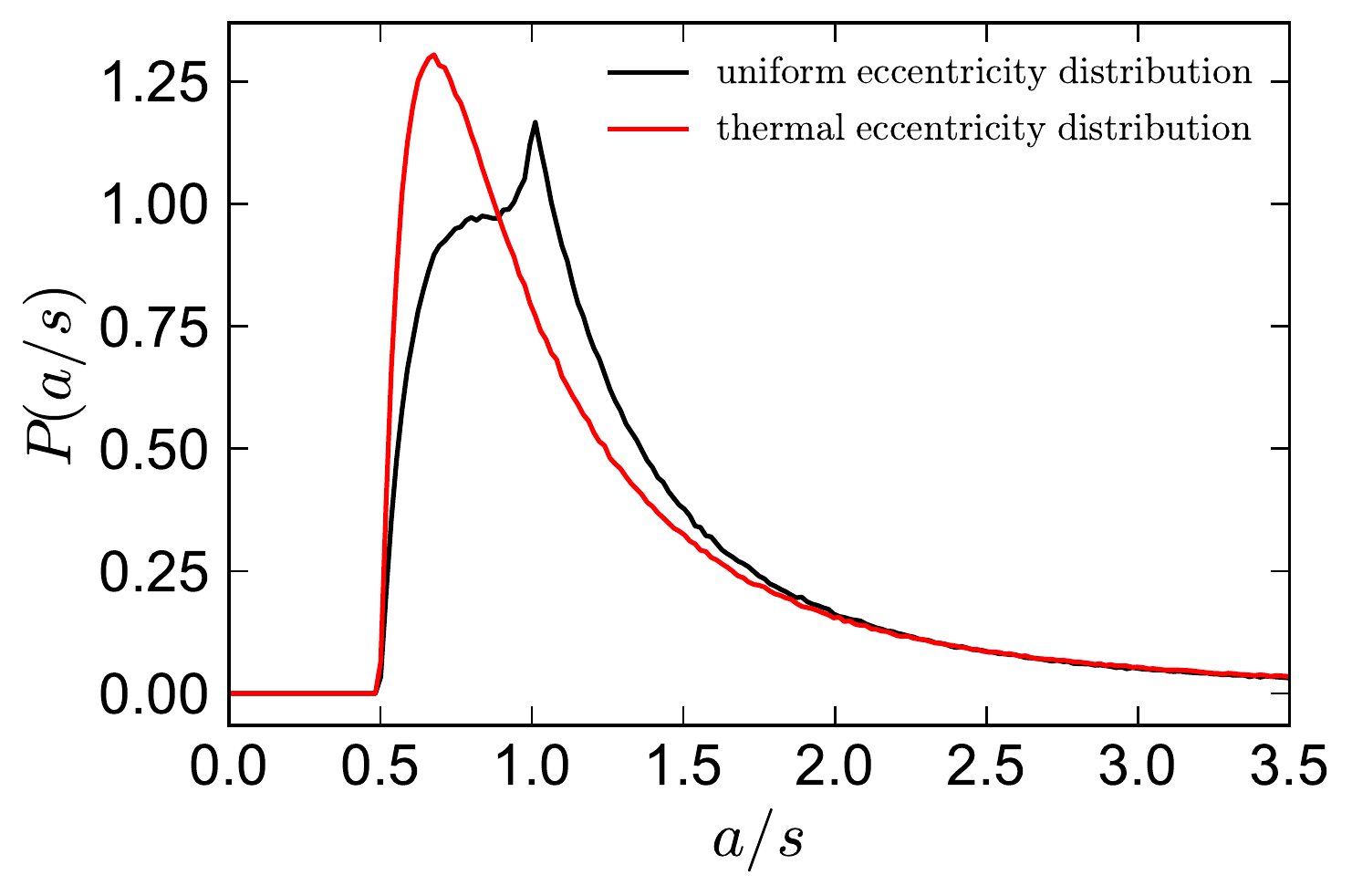}
\caption{Distribution of the conversion factor, $P(a/s)$, from projected separation $s$ to semi-major axis $a$ for randomly oriented orbits. We compare distributions for a binary population with a uniform eccentricity distribution (as is assumed in our population synthesis simulations; black) and a thermal eccentricity distribution ($f\left(e\right)\,{\rm d}e=2e\,{\rm d}e$, red).}
\label{fig:rand_orbits}
\end{figure}

When comparing \texttt{BSE} simulations to our observational constraints, it is necessary to translate between distributions of the intrinsic semi-major axis, $a$, and the projected separation, $s$. To determine an appropriate conversion factor distribution, we mock-observe an ensemble of $3\times 10^6$ simulated binaries. 

Each binary is observed along a random line-of-sight (resulting in an a $p(i)\,{\rm d}i = \sin(i)\,{\rm d}i$ inclination distribution) at a random time in its orbit. We consider two eccentricity distributions: a uniform distribution, which is roughly consistent with observations \citep[e.g.][]{Duchene_2013, Tokovinin_2016} and is also assumed in our \texttt{BSE} simulations, and a thermal distribution, $p(e)\,{\rm d}e = 2e\,{\rm d}e$, which is predicted if the phase-space distribution of binaries is a function of energy alone \citep[e.g.][]{Jeans_1919}. The distributions of stellar mass, mass ratio, and orbital period have no effect on the conversion factor between $s$ and $a$. 

The results of these simulations are shown in Figure~\ref{fig:rand_orbits}; they are identical to the findings of \citet{Dupuy_2011}, who performed a similar experiment. For each binary simulated in our population synthesis experiments in Section~\ref{sec:theory}, we multiply the semimajor axis by a random draw from the $P(a/s)$ distribution for a uniform eccentricity distribution in order to construct the $P(s)$ distribution. For both eccentricity distributions, $s$ and $a$ differ by less than a factor of two for a large majority of binaries. Because we consider separation distributions over separations varying by many orders of magnitude, $P(s)$ and $P(a)$ are nearly indistinguishable.

\section{Distance dependence}
\label{sec:near_vs_far}

\begin{figure}
\includegraphics[width=\columnwidth]{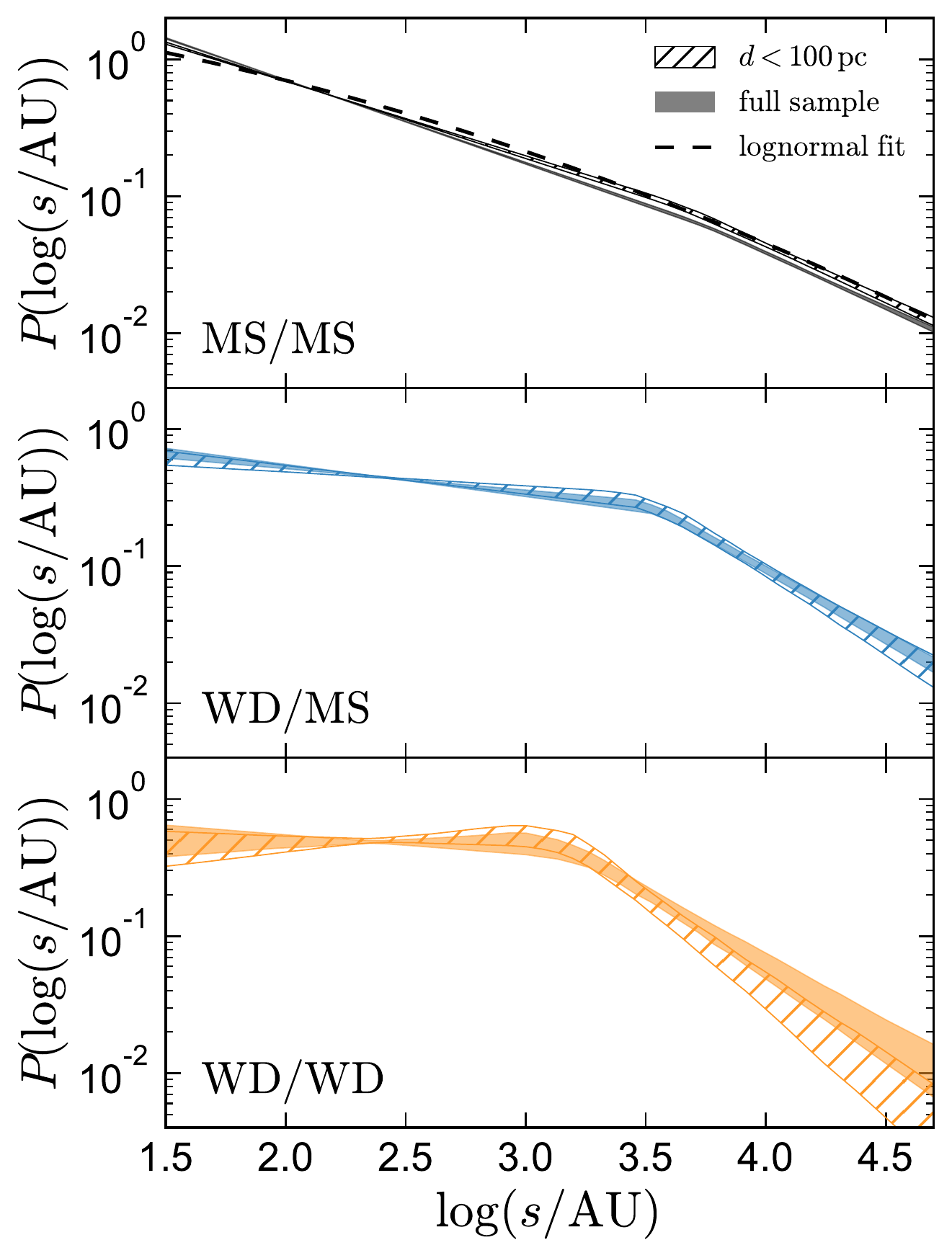}
\caption{$1\sigma$ constraints on the separation distributions of different classes of wide binaries. We compare constrains obtained from fitting the full sample within 200 pc (solid; identical to Figure~\ref{fig:corner_fits}) to those obtained when fitting only the nearest objects within 100 pc (hatched); they are essentially consistent. Distance distributions are shown in Figure~\ref{fig:summary}. In the top panel, we also show the best-fit lognormal distribution for the full MS/MS sample.}
\label{fig:near_far}
\end{figure}

In Figure~\ref{fig:near_far}, we compare the separation distributions obtained for our full binary sample to those obtained when we apply our inference only to the binaries within 100 pc (i.e., $\varpi > 10$\,mas). We expect the intrinsic separation distribution of the 100 pc sample to be similar to that of the full sample. Because the typical parallax and proper motion uncertainties, angular separation, and expected contamination rate of the 100 pc sample are different from those of the full catalog, this experiment provides a useful test of the inference described in Section~\ref{sec:inference}. 

The 100 pc sample contains 10179, 977, and 177 MS/MS, WD/MS, and WD/WD binaries, respectively. Figure~\ref{fig:near_far} shows that the constraints obtained for each class of binary are quite similar in the 100 pc sample to in the full sample. Although there are slight differences in the constraints, the separation distributions are consistent within 2$\sigma$ at all separations, and none of our conclusions would change if we were to fit only the 100 pc sample. 

In the top panel of Figure~\ref{fig:near_far}, we also show the best-fit distribution obtained for MS/MS binaries when we replace the default broken power law parameterization of $\phi(s)$ with a lognormal; i.e., a Gaussian in $\log(s)$. The formal best-fit parameters are $\mu_{\log s/{\rm AU}}=-0.93$ and $\sigma_{{\rm log}s/{\rm AU}}=1.69$. However, we caution that because there is no strong evidence of a turnover in $\phi(s)$ over the range of separations probed by our data, $\mu_{\log s}$ and $\sigma_{\log s}$ are strongly degenerate, and the best-fit value of $\mu_{\log s}$ is meaningless \citep[see][their Figure 1]{ElBadry_2017}. For example, a lognormal with $\mu_{\log s/{\rm AU}}=1.0$ and $\sigma_{{\rm log}s/{\rm AU}}=1.26$ provides a comparably good fit.

\section{ADQL query}
\label{sec:query}
The ADQL query used to obtain our initial sample of candidate binaries (before removing clusters, triples, etc., and before applying all the quality cuts listed in Section~\ref{sec:quality}) is reproduced below. The search approach is modeled after that used by \citet{Fouesneau_2018}. We follow the variable naming convention from the {\it Gaia} Archive.

\begin{verbatim}
SELECT g2.source_id as source_id2, 
    g2.ra as ra2, g2.dec as dec2, 
    g2.astrometric_chi2_al as 
    astrometric_chi2_al_2, 
    g2.astrometric_n_good_obs_al 
    as astrometric_n_good_obs_al2, 
    g2.phot_g_mean_flux_over_error 
    as phot_g_mean_flux_over_error2, 
    g2.phot_rp_mean_flux_over_error 
    as phot_rp_mean_flux_over_error2, 
    g2.phot_bp_mean_flux_over_error 
    as phot_bp_mean_flux_over_error2, 
    g2.phot_bp_rp_excess_factor 
    as phot_bp_rp_excess_factor2,
    g2.pmra as pmra2, g2.pmra_error 
    as pmra_error2, g2.pmdec as pmdec2,
    g2.pmdec_error as pmdec_error2,
    g2.phot_g_mean_mag as 
    phot_g_mean_mag2, g2.phot_bp_mean_mag 
    as phot_bp_mean_mag2, 
    g2.phot_rp_mean_mag as 
    phot_rp_mean_mag2, g2.parallax as 
    parallax2, g2.parallax_over_error as 
    parallax_over_error2, g2.radial_velocity 
    as radial_velocity2, 
    g2.radial_velocity_error as 
    radial_velocity_error2, g2.rv_nb_transits 
    as rv_nb_transits2, t1.source_id, t1.ra, 
    t1.dec, t1.pmra, t1.pmra_error, t1.pmdec, 
    t1.pmdec_error, t1.phot_g_mean_mag, 
    t1.phot_bp_mean_mag, t1.phot_rp_mean_mag, 
    t1.parallax, t1.parallax_over_error, 
    t1.astrometric_chi2_al, 
    t1.astrometric_n_good_obs_al, 
    t1.phot_g_mean_flux_over_error, 
    t1.phot_rp_mean_flux_over_error, 
    t1.phot_bp_mean_flux_over_error, 
    t1.phot_bp_rp_excess_factor, 
    t1.radial_velocity, t1.radial_velocity_error, 
    t1.rv_nb_transits, 
    distance(POINT('ICRS', t1.ra, t1.dec), 
    POINT('ICRS', g2.ra, g2.dec)) AS pairdistance
FROM (select * from gaiadr2.gaia_source where 
    parallax between 5 and 1000 and 
    parallax_over_error > 20 and 
    phot_g_mean_flux_over_error > 50 and 
    phot_rp_mean_flux_over_error > 20 and 
    phot_bp_mean_flux_over_error > 20) as t1, 
    (select * from gaiadr2.gaia_source 
    where bp_rp is not null and 
    parallax_over_error > 5 and 
    phot_g_mean_flux_over_error > 50 and 
    phot_rp_mean_flux_over_error > 10 and 
    phot_bp_mean_flux_over_error > 10) as g2 
WHERE 1 = contains(POINT('ICRS', g2.ra, g2.dec), 
    CIRCLE('ICRS', t1.ra, t1.dec, 
    1.4e-2*t1.parallax)) 
AND t1.source_id != g2.source_id 
AND abs(1/t1.parallax - 1/g2.parallax) - 
    2*0.01745*distance(POINT('ICRS', t1.ra, 
    t1.dec), POINT('ICRS', g2.ra, g2.dec))/
    t1.parallax < 3*sqrt(power(t1.parallax_error, 
    2)/power(t1.parallax, 4) + power(
    g2.parallax_error, 2)/power(g2.parallax, 4))
AND sqrt(power((t1.pmra - g2.pmra), 2) + 
    power((t1.pmdec - g2.pmdec), 2)) - 
    (7.42e-3 * power(t1.parallax, 1.5) * 
    power(distance(POINT('ICRS', t1.ra, t1.dec),
    POINT('ICRS', g2.ra, g2.dec)), -0.5)) < 
    3*sqrt(((power((t1.pmra_error), 2) + power(
    (g2.pmra_error), 2)) * power((t1.pmra - 
    g2.pmra), 2) + (power((t1.pmdec_error), 2) + 
    power((g2.pmdec_error), 2)) * power(
    (t1.pmdec - g2.pmdec), 2))/(power(
    (t1.pmra - g2.pmra), 2) + power(
    (t1.pmdec - g2.pmdec), 2)))
AND sqrt(((power((t1.pmra_error), 2) + 
    power((g2.pmra_error), 2)) * power(
    (t1.pmra - g2.pmra), 2) + (power(
    (t1.pmdec_error), 2) + power(
    (g2.pmdec_error), 2)) * power((t1.pmdec -
    g2.pmdec), 2))/(power((t1.pmra - g2.pmra), 
    2) +  power((t1.pmdec - g2.pmdec), 2))) < 1.5
\end{verbatim}

In practice, this query will not finish within the default time limits on the {\it Gaia} Archive. We therefore split it on source id into many smaller queries. That is, we add the condition \texttt{where source\_id between A and B}, where A and B are integers, to both clauses in the \texttt{FROM} block. We find splitting on healpix level 2 (i.e., 192 regions of area $\approx 215\,\rm degree^2$) to work well. Splitting up the sky during the search could in principle introduce a bias against pairs with large separations, which are more likely to be missed because the two component stars fall in different regions of the sky. We have verified through simulations that any resulting bias is at the $<3\%$ level, because the typical angular separations of even the widest binaries in our catalog are small compared to the size of each search region. 

\section{Catalog}
\label{sec:catalog}
Table~\ref{tab:catalog} provides a summary of our catalog of wide binaries within 200 pc. A full version of the catalog is available online. 

\begin{table*}
\centering
\caption{Catalog description}
\label{tab:catalog}
\begin{tabular}{p{5.0cm} | p{1.5cm}| p{10cm} } 

Column & units & Description \\
\hline
\texttt{astrometric\_chi2\_al}               &         & astrometric goodness-of-fit ($\chi^2$) in the along-scan direction; star 1   \\
\texttt{astrometric\_chi2\_al\_2}            &         & astrometric goodness-of-fit ($\chi^2$) in the along-scan direction; star 2  \\
\texttt{astrometric\_n\_good\_obs\_al}       &         & number of good CCD transits; star 1 \\
\texttt{astrometric\_n\_good\_obs\_al2}      &         & number of good CCD transits; star 2  \\
\texttt{binary\_class}                       &         & type of binary; \texttt{MSMS}, \texttt{WDMS}, or \texttt{WDWD}  \\
\texttt{dec}                                 & deg     & declination; star 1  \\
\texttt{dec2}                                & deg     & declination; star 2  \\
\texttt{pairdistance}                        & deg     & angular separation between star 1 and star 2  \\
\texttt{parallax}                            & mas     & parallax; star 1  \\
\texttt{parallax2}                           & mas     & parallax; star 2  \\
\texttt{parallax\_over\_error}               &         & parallax divided by its error; star 1  \\
\texttt{parallax\_over\_error2}              &         & parallax divided by its error; star 2 \\
\texttt{phot\_bp\_mean\_flux\_over\_error}   &         & integrated BP mean flux divided by its error; star 1  \\
\texttt{phot\_bp\_mean\_flux\_over\_error2}  &         & integrated BP mean flux divided by its error; star 2  \\
\texttt{phot\_bp\_mean\_mag}                 & mag     & integrated BP mean magnitude; star 1  \\
\texttt{phot\_bp\_mean\_mag2}                & mag     & integrated BP mean magnitude; star 2  \\
\texttt{phot\_bp\_rp\_excess\_factor}        &         & ratio of total integrated BP and RP flux to G-band flux; star 1 \\
\texttt{phot\_bp\_rp\_excess\_factor2}       &         & ratio of total integrated BP and RP flux to G-band flux; star 2  \\
\texttt{phot\_g\_mean\_flux\_over\_error}    &         & integrated G-band mean flux divide by its error; star 1  \\
\texttt{phot\_g\_mean\_flux\_over\_error2}   &         & integrated G-band mean flux divide by its error; star 2  \\
\texttt{phot\_g\_mean\_mag}                  & mag     & G-band mean magnitude (Vega scale); star 1  \\
\texttt{phot\_g\_mean\_mag2}                 & mag     & G-band mean magnitude (Vega scale); star 2  \\
\texttt{phot\_rp\_mean\_flux\_over\_error}   &         & integrated RP mean flux divided by its error; star 1  \\
\texttt{phot\_rp\_mean\_flux\_over\_error2}  &         & integrated RP mean flux divided by its error; star 2  \\
\texttt{phot\_rp\_mean\_mag}                 & mag     & integrated RP mean magnitude; star 1  \\
\texttt{phot\_rp\_mean\_mag2}                & mag     & integrated RP mean magnitude; star 2  \\
\texttt{pmdec}                               & mas\,yr$^{-1}$ & proper motion in the declination direction; star 1  \\
\texttt{pmdec2}                              & mas\,yr$^{-1}$ & proper motion in the declination direction; star 2  \\
\texttt{pmdec\_error}                        & mas\,yr$^{-1}$ & standard error of proper motion in the declination direction; star 1  \\
\texttt{pmdec\_error2}                       & mas\,yr$^{-1}$ & standard error of proper motion in the declination direction; star 2  \\
\texttt{pmra}                                & mas\,yr$^{-1}$ & proper motion in right ascension direction; i.e., $\mu_{\alpha}^{*} = \mu_{\alpha}\cos{\delta}$; star 1  \\
\texttt{pmra2}                               & mas\,yr$^{-1}$ & proper motion in right ascension direction; i.e., $\mu_{\alpha}^{*} = \mu_{\alpha}\cos{\delta}$; star 2  \\
\texttt{pmra\_error}                         & mas\,yr$^{-1}$ & standard error of proper motion in right ascension direction; star 1  \\
\texttt{pmra\_error2}                        & mas\,yr$^{-1}$ & standard error of proper motion in right ascension direction; star 2  \\
\texttt{ra}                                  & deg     & right ascension; star 1  \\
\texttt{ra2}                                 & deg     & right ascension; star 2  \\
\texttt{radial\_velocity}                    & km\,s$^{-1}$ & spectroscopic barycentric radial velocity; star 1  \\
\texttt{radial\_velocity2}                   & km\,s$^{-1}$ & spectroscopic barycentric radial velocity; star 1  \\
\texttt{radial\_velocity\_error}             & km\,s$^{-1}$ & standard error of spectroscopic barycentric radial velocity; star 1  \\
\texttt{radial\_velocity\_error2}            & km\,s$^{-1}$ & standard error of spectroscopic barycentric radial velocity; star 2  \\
\texttt{rv\_nb\_transits}                    &         & number of transits used to compute radial velocity; star 1  \\
\texttt{rv\_nb\_transits2}                   &         & number of transits used to compute radial velocity; star 2  \\
\texttt{s\_AU}                               & AU      & projected physical separation between two stars \\
\texttt{source\_id}                          &         & {\it Gaia} source id (int64); star 1  \\
\texttt{source\_id2}                         &         & {\it Gaia} source id (int64); star 2  \\
\hline
\end{tabular}
\begin{flushleft}
{\bf Note}: Each row in the catalog corresponds to a single binary; ``star 1'' and ``star 2'' designations in each binary are arbitrary. Full descriptions of {\it Gaia} variables can be found at \Gaiaurl.
\end{flushleft}
\end{table*}

\bsp	
\label{lastpage}
\end{document}